\documentclass[sigconf]{acmart}
\AtBeginDocument{%
  \providecommand\BibTeX{{%
    \normalfont B\kern-0.5em{\scshape i\kern-0.25em b}\kern-0.8em\TeX}}}
\usepackage{colortbl}
\usepackage{array}
\usepackage{caption}
\usepackage{subcaption}
\usepackage{geometry}
\usepackage{xcolor}
\usepackage{booktabs}
\usepackage{multirow}
\usepackage{tabularx}
\usepackage{graphicx}
\usepackage{longtable}
\usepackage{appendix}
\usepackage{url}
\usepackage{hyperref}
\hypersetup{
    colorlinks=false,
    hidelinks
}
\begin{document}

\title[Relief or displacement? How teachers are negotiating generative AI's role in their professional practice]%
{Relief or displacement? How teachers are negotiating generative AI's role in their professional practice
}

\renewcommand{\shortauthors}{Dangol et al.}

\author{Aayushi Dangol}
\affiliation{%
  \institution{University of Washington}
  \city{Seattle}
  \state{WA}
  \country{USA}
}

\author{Smriti Kotiyal}
\affiliation{%
  \institution{University of Washington}
  \city{Seattle}
  \state{WA}
  \country{USA}
}

\author{Robert Wolfe}
\affiliation{%
  \institution{Rutgers University}
  \city{New Brunswick}
  \state{NJ}
  \country{USA}
}

\author{Alex J. Bowers}
\affiliation{%
  \institution{Columbia University}
  \city{New York}
  \state{NY}
  \country{USA}
}

\author{Antonio Vigil}
\affiliation{%
  \institution{Aurora Public Schools}
  \city{Aurora}
  \state{CO}
  \country{USA}
}

\author{Jason Yip}
\affiliation{%
  \institution{University of Washington}
  \city{Seattle}
  \state{WA}
  \country{USA}
}

\author{Julie A. Kientz}
\affiliation{%
  \institution{University of Washington}
  \city{Seattle}
  \state{WA}
  \country{USA}
}

\author{Suleman Shahid}
\affiliation{%
  \institution{Lahore University}
  \city{Lahore}
  \state{Punjab}
  \country{Pakistan}
}

\author{Tom Yeh}
\affiliation{%
  \institution{University of Colorado Boulder}
  \city{Boulder}
  \state{CO}
  \country{USA}
}

\author{Vincent Cho}
\affiliation{%
  \institution{Boston College}
  \city{Chestnut Hill}
  \state{MA}
  \country{USA}
}

\author{Katie Davis}
\affiliation{%
\institution{University of Washington}
  \city{Seattle}
  \state{WA}
  \country{USA}
}

\renewcommand{\shortauthors}{Dangol et al.}

\begin{abstract}
As generative AI (genAI) rapidly enters classrooms, accompanied by district-level policy rollouts and industry-led teacher trainings, it is important to rethink the canonical “adopt and train” playbook. Decades of educational technology research show that tools promising personalization and access often deepen inequities due to uneven resources, training, and institutional support. Against this backdrop, we conducted semi-structured interviews with 22 teachers from a large U.S. school district that was an early adopter of genAI. Our findings reveal the motivations driving adoption, the factors underlying resistance, and the boundaries teachers negotiate to align genAI use with their values. We further contribute by unpacking the sociotechnical dynamics---including district policies, professional norms, and relational commitments---that shape how teachers navigate the promises and risks of these tools.
\end{abstract}


\maketitle

\section{Introduction}
Visions for how generative AI (genAI) could reshape education are rapidly evolving, fueled by billions of dollars in investment from companies like OpenAI, Microsoft, and Google, alongside federal initiatives that position genAI integration as a national priority \cite{nyt2025microsoft, Trump2025AdvancingAIEducation}. A recent U.S. executive order, for example, directs schools to incorporate genAI “across all subject areas” and train teachers to develop student expertise “from an early age” \citep{Trump2025AdvancingAIEducation}. Moreover, whether formally or informally, genAI platforms have already reshaped the ecology of schooling \citep{cukurova2024ai, office2023artificial, lin2024s}. Students are using genAI to plan, write, and revise essays, design presentations, and produce creative media, while school districts are beginning to formalize adoption. One sign of this institutional shift is the growing number of large-scale public–private partnerships aimed at accelerating teacher readiness for genAI integration \citep{nyt2025miamidade, nyt2025bankroll}. For example, the American Federation of Teachers, one of the nation’s largest teachers’ unions, recently announced a \$23 million partnership with major genAI industry leaders to train approximately 400,000 K–12 educators on genAI instruction. 

Though these developments seem promising, successful teaching is relational, and it entails more than simply conveying the right information at the right time. What's more, the introduction of new technologies into classrooms can have unintended and sometimes adverse consequences, as in the case of smartphones, or even serve to widen and reinforce inequities they were expected to rectify \citep{rafalow2020digital, toyama2015geek, reich2012state, reich2017good}. If, as is often the case, policy and purchasing decisions are made at the state or district level with insufficient input from teachers \citep{williamson2024ai} or regard for how the new technology will be integrated into existing practices, genAI has little chance of standing up to the realities of classroom instruction.

All of this raises the question, then, of how teachers understand and use genAI within the broader context of its systems-level introduction into schools and school districts. Early studies suggest that teachers use the technology to develop assignments and lessons; provide feedback on student work; engage with multilingual families; and track student progress \citep{Cipolla2024GenerativeAI, Goldstein2025TeachersAI}. These practices are closely tied to efficiency gains, as teachers report reallocating saved hours toward individualized feedback, differentiated lesson design, and student support. At the same time, teachers' use of genAI is often informal and experimental, involving new forms of educational labor such as refining prompts, verifying outputs, and negotiating concerns about fairness, authorship, and bias {\citep{GreeneNolanPhD2024TeachingPG, Biton2025LearningTC, Elsayary2025ExaminingTR}.

Because teachers sit at the intersection of classroom practice and broader policy shifts, bringing their perspectives into conversations about genAI integration is critical for shaping equitable, effective, and contextually grounded decisions. While much of the existing literature on genAI in education centers on students’ use of these tools \citep{ammari2025students,zheng2025students,schneider2025thematic,Lee2024CheatingIT, pu2025can}, our study foregrounds teaching practice, examining how teachers engage with and adapt their instructional approaches in response to genAI. In doing so, we recognize that while our work has direct implications for student learning, it does so through its emphasis on teacher decision-making. This leads us to ask:
\begin{itemize}
\item[] \textbf{RQ1:} How do teachers perceive the introduction of genAI tools in K–12 settings?
\item[] \textbf{RQ2:} What opportunities and challenges arise as schools introduce genAI tools into teaching processes?
\end{itemize}

To answer these questions, we conducted semi-structured interviews with 22 teachers from BPS (anonymized for review), a large US public school district. Serving students who collectively speak more than 160 languages, BPS has long faced challenges related to supporting a highly multilingual population, ongoing teacher shortages, and heavy instructional workloads. In late 2023, the district encouraged educators to integrate MagicSchoolAI--a teacher-focused genAI platform--into their classrooms as part of a broader effort to address these pressures. 

Findings from our thematic analysis of the interview transcripts revealed that teachers are not simply adopting or resisting genAI but are actively negotiating its role within their professional identities, pedagogical goals, and classroom relationships. Central to this negotiation is a tension between relief and displacement. While teachers welcomed genAI’s ability to streamline routine tasks, alleviate burnout, and expand instructional possibilities, they also worried about what might be lost when intellectual and relational aspects of their work were delegated to genAI systems. By surfacing how teachers navigate this tension and set boundaries around their practice, our study contributes a teacher-centered, sociotechnical account of genAI integration in K–12 education, highlighting the conditions necessary for responsible and sustainable adoption. Specifically, we show how institutional policies, peer networks, and professional norms shape teachers’ decisions and offer implications for the design of genAI-supported tools that balance technological affordances with teachers’ pedagogical commitments and practical realities.

In the remainder of this paper, we begin by reviewing related work on genAI in education. We then describe our study’s methodological approach, including data collection procedures and analysis. Finally, we present our findings and discuss their implications for designing genAI-supported tools that align with teachers’ values and reflect the complexities of real-world classrooms.

\section{Related Work}
In this section, we review three strands of scholarship that inform our study. First, we examine emerging work on the use of genAI in teaching and learning, highlighting both creative possibilities and teacher-facing practices. Second, we situate these developments within the broader history of educational technology. Finally, we turn to the Teacher Response Model (TRM) \citep{kopcha2020process}, which provides a process-oriented lens on technology integration, foregrounding teachers’ sensemaking, values, and in-the-moment decision-making as central to understanding how genAI is taken up in classrooms.

\subsection{Generative AI for Teaching and Learning}
Recent advances in genAI have introduced new possibilities for teaching and learning, particularly through large language models (LLMs) and text-to-image models \citep{Laato2023AIAssistedLW, Bonner2023LARGELM, bewersdorff2025taking}. HCI research has examined the potential of genAI in education across several domains, including creative authoring and design support (e.g., text-to-image systems that enable students to illustrate stories or concepts) \citep{Maceli2024IncorporatingUU, Wang2025ScaffoldingCI, he2025carlitos}, collaborative writing and knowledge construction (e.g., LLM-powered tools that scaffold brainstorming, drafting, and revision) \citep{Ye2025TheDS, ravi2025co, kang2025tutorcraftease}, and dialog-based tutoring systems that allow open-ended question–answer exchanges \citep{Banjade2024EmpoweringEB, calo2024towards, Hu2025GenerativeAI}. Building on these broader explorations of creative and collaborative uses, researchers have also turned their attention to core instructional practices where genAI might augment teachers’ capacities \citep{acevedo2025exploring, lu2024supporting, macdowell2024preparing}. 

One area of growing focus is assessment. Emerging research shows that teachers are using genAI both to generate practice materials and to provide formative feedback \citep{Maity2024TheFO, koh2023learning, moorhouse2023generative}. For example, LLMs have been applied to produce customized practice questions in subjects such as mathematics \citep{Yen2023ThreeQC, Li2024AutomateKC}, English language learning \citep{tafazoli2024exploring, kohnke2024technostress}, and data science \citep{shen2024implications}. In parallel, some educators have leveraged AI-generated explanations to help programming students iterate on their assignments more quickly \citep{ghimire2024coding}, while science instructors have experimented with models that classify short answers in ways that approximate expert judgment \citep{Impey2024UsingLL}. Beyond assessment, teachers are increasingly adopting genAI for instructional design \citep{Kong2024AHL, ellis2023new}. Studies document its use in drafting lesson plans \citep{laak2024generative, ng2025opportunities}, producing differentiated practice materials, and adapting activities for diverse student needs \citep{lewis2025exploring, suh2024opportunities}. GenAI has also been piloted in interactive classroom contexts, where genAI agents act as peer teammates, coaches, or role-play partners in collaborative problem-solving \citep{mollick2024ai, song2024students}. 

While these studies illustrate how individual teachers are beginning to integrate genAI into their instructional practices, large-scale implementation efforts provide a complementary perspective on how such tools function in practice. For example, the Indiana Department of Education’s 2023–2024 pilot across five AI platforms illustrates both the promise and fragility of large-scale integration \citep{IndianaDOE2024AIReport}. Teachers highlighted time-saving benefits and opportunities for personalized learning, yet many also reported difficulties with usability, uneven subject coverage, and a lack of training support \citep{IndianaDOE2024AIReport}. These patterns echo national trends, where teachers increasingly express curiosity about genAI but often lack clear guidance, professional development, and policy frameworks to support meaningful use in classrooms \citep{Joshi2025StrategicIO, Kim2025PerceptionsAP, Ghimire2024FromGT}. Students, too, are experiencing these inconsistencies as access, expectations, and instructional practices differ widely across classrooms and school districts \citep{pu2025can, baidoo2023education}. At the same time, long-standing inequities in resources, AI literacy, and prior experience with AI technologies raise critical questions about who benefits from these tools and on what terms. In the next section, we situate these emerging uses of genAI within the broader history of educational technologies, highlighting how prior waves of innovation reveal patterns that can help interpret the promises and challenges of the current moment.

\subsection{Placing Generative AI in the Context of Prior Educational Technologies}
The introduction of genAI in schools builds on a long history of educational technologies, from early computer-assisted instruction \citep{atkinson1968computer, blok2002computer} and intelligent tutoring systems \citep{graesser2012intelligent, anderson1985intelligent} to MOOCs \citep{wang2015content, baturay2015overview} and adaptive learning platforms \citep{kem2022personalised, divanji2023one}. Each wave has carried the promise of personalized instruction, increased engagement, and expanded access. Prior research, however, shows that such promises are unevenly realized as technologies often deliver the greatest benefits where schools have the resources to integrate them thoughtfully \citep{reich2017good, rafalow2020digital}. Well-resourced schools, for example, can provide teacher training and time for experimentation, allowing new tools to support more meaningful and interactive learning experiences \citep{reich2020failure, delgado2015educational, Johnson2016ChallengesAS, Selwyn2013EmpoweringTW}. In contrast, schools with fewer resources may incorporate new technologies in ways that largely replicate existing practices by streamlining content delivery or administrative tasks rather than enabling new forms of pedagogy \citep{reich2017good, toyama2015geek, cuban2001oversold, cuban2015dubious}.

Scholars describe this pattern as one of amplification, where technologies magnify the strengths and limitations of the contexts into which they are introduced \citep{toyama2015geek}. For example, Reich et al.’s \citep{reich2012state} analysis of over 180,000 public school wikis found that teachers in affluent schools used often wikis for student-driven collaboration, while under-resourced schools employed them primarily as static places to store information. Similarly, Rafalow’s \citep{rafalow2020digital} ethnography of different middle schools described how teachers in a predominantly white, elite private school encouraged students to interact with technologies in open-ended and creative ways, while those in a Latinx-majority school discouraged digital “play” and prioritized rote learning. These examples suggest that the impact of educational technology is rarely determined by the tool itself \citep{Selwyn2013EmpoweringTW, Schuck2008ClassroomBasedUO}. Instead, outcomes depend on how technologies are interpreted, implemented, and adapted by educators within specific institutional, cultural, and material contexts. 

In light of this history, the current wave of genAI adoption raises important questions. While genAI powered educational technologies and LLMs like ChatGPT, Gemini, and Claude are being rapidly introduced into K–12 settings, little is known about how teachers are using these tools, how they reconcile them with pedagogical goals, or how they navigate tensions between efficiency with the need for accuracy and the responsibility to support meaningful learning experiences. Our work builds on this prior insight by shifting attention from the rhetoric of genAI-enabled transformation to the practical, situated work of educators as they encounter and integrate genAI into their teaching practice.

\subsection{Teacher Response Model as a Lens for Understanding Generative AI Integration}
To understand how K-12 teachers are integrating genAI into their instructional practices, we draw on a sociotechnical systems approach to genAI and teacher practice that recognizes the mutual constitution of technological and social dimensions in educational contexts \citep{kaghan2001out, latour1987science, leonardi2009crossing}. This perspective acknowledges that focusing solely on genAI's technical capabilities without attending to how these tools are incorporated into specific cultural and institutional contexts risks replicating inequitable patterns of EdTech adoption \citep{ames2015charismatic, Sommerfeld2020AmesMG}. Within this broader tradition, we specifically adapt the Teacher Response Model (TRM) introduced by Kopcha et al. \citep{kopcha2020process}, which reconceptualizes technology integration not as a fixed outcome, but as a fluid, situated decision-making process embedded in teachers’ everyday work. This includes how they perceive the tool’s possibilities, how they respond to emergent needs in the moment, and how internal factors (e.g., beliefs, experience, professional identity) interact with external conditions (e.g., curricular goals, student behavior, resource access) in real time \citep{kopcha2020process}. This process-oriented lens is particularly well-suited to the case of genAI. Unlike more bounded educational technologies like cognitive tutors or adaptive learning platforms, genAI systems function through open-ended interaction, requiring teachers to actively interpret what the system can do, when it is appropriate to use, and how its outputs align or conflict with their pedagogical intentions. As a result, genAI amplifies the interpretive, relational, and ethical dimensions of teaching, placing new demands on educators’ professional judgment and in-the-moment decision-making. Below, we distill elements of TRM into three key characteristics that inform our study.

\renewcommand{\theenumi}{T\arabic{enumi}}
\begin{enumerate}
\item \textbf{Technology integration is value-driven:} Teachers use technology not to achieve externally defined standards of “high-level use,” but to accomplish what they perceive as best for their students and their teaching context \citep{kopcha2020process}. These values may reflect instructional goals, social-emotional needs, time constraints, or classroom management concerns. In the case of genAI, this means examining how teachers define ``helpful'' or ``harmful'' use in relation to their professional and ethical commitments.

\item \textbf{Decision-making is embedded in a dynamic system:} Classrooms are complex, ever-changing environments in which teachers constantly adapt to shifting needs, constraints, and feedback loops \citep{kopcha2020process}. Technology integration decisions are thus emergent and context-dependent. As teachers experiment with genAI, we attend to how these systems are taken up, resisted, or reconfigured in response to unexpected student reactions, institutional mandates, or evolving pedagogical goals.

\item \textbf{Teacher action is shaped by perceptions of what is possible:} Drawing on embodied and situated cognition \citep{Schoenfeld2015HowWT, Alibali2018EmbodiedCI}, the TRM emphasizes that teachers act based on what they perceive as doable within a given moment \citep{kopcha2020process}. These perceptions are informed by prior experience, access, familiarity, and perceived risk. With genAI, what is perceived as “possible” is often in flux, making teacher experimentation, improvisation, and uncertainty central to the integration process.
\end{enumerate} 

Drawing on this theoretical framing, our study examines how teachers navigate the situated, interpretive process of integrating genAI into their pedagogy. In doing so, we position teachers’ perspectives as a useful lens for understanding how professional values, contextual constraints, and sociotechnical dynamics shape both their practices and their evolving perceptions of what genAI can and should do.

\section{Methods}
Our study was conducted at BPS (anonymized for review), a large urban-suburban school district located in the United States. As of the 2024–25 school year, BPS serves more than 38,000 students across 59 schools, including traditional elementary, middle, and high schools, as well as charter and magnet programs. The district serves a multilingual and socioeconomically diverse student population, with over 160 languages spoken and 74.6\% of students eligible for free or reduced-price lunch. Since early 2023, BPS has actively encouraged teachers to explore genAI tools in their instructional practice, including the district-wide adoption of MagicSchool AI,\footnote{\url{https://www.magicschool.ai}} an educator-facing platform that provides AI-generated support.  While some BPS educators had attended workshops or webinars focused on educational genAI tools, others had independently experimented with various genAI tools. Interviews were conducted during spring 2025, at a time when no formal, district-wide policies for classroom genAI use had yet been established. This allowed us to examine how teachers were independently navigating emerging tools in the absence of centralized guidance.

\subsection{Semi-Structured Interviews}
\subsubsection{Participants}
We recruited participants in collaboration with the Director of Educational Technology at BPS, who distributed a recruitment email to teachers across the district. Interested teachers completed an initial screening survey that collected information about their teaching context, grade level, and prior experience with genAI tools. Based on these responses, we conducted purposive sampling \citep{Palinkas2015PurposefulSF} to select a diverse group of educators (see Table \ref{tab:teacher_participant_roles}). Participants taught across elementary, middle, and high school grade levels and represented a range of subject areas, including STEM, Social Studies, Special Education, Culturally \& Linguistically Diverse Education (CLDE). Teaching experience ranged from 1–2 years to over 11 years, with 63.6\% of teachers reporting over a decade of classroom experience. Time in their current role also varied, with 40.9\% participants in their first 1–2 years, 31.8\% had 3–5 years of experience, and 27.3\% had been in the same position for over 6 years. Participants also reported using a wide range of genAI tools, with most educators experimenting with multiple platforms. Common tools included ChatGPT, MagicSchool AI, Canva, TalkingPoints, Diffit, and Gemini.

\subsubsection{Procedure}
Each interview lasted approximately one hour and was conducted remotely via Zoom. We developed our interview protocol based on our research questions and theoretical framing, drawing on the TRM \citep{kopcha2020process}. During the interviews, we asked teachers to describe how they first encountered genAI tools, how they currently use them in their professional practice, and the broader shifts they have observed in their schools. Teachers also reflected on both opportunities and challenges of using genAI tools in their classrooms. The semi-structured online format allowed teachers to participate from their own homes and enabled the research team to probe further or clarify ideas depending on each participant’s experience. Teachers received a \$75 Amazon gift card for their participation.

\subsubsection{Data Analysis}
We audio recorded all interviews and transcribed them for analysis using Rev.ai,\footnote{\url{https://www.rev.ai/}} a secure transcription service. We analyzed the teacher interview data using Reflexive Thematic Analysis \citep{Braun2020OneSF, Braun2019ReflectingOR}, guided by our research questions and theoretical framing. In particular, the TRM \citep{kopcha2020process} shaped our analytical lens by sensitizing us to teachers’ decision-making about genAI integration, including how they perceived the tool’s possibilities, navigated contextual constraints, and balanced evolving pedagogical goals.

The coding team, comprising the first and second authors, began by independently open coding two randomly selected transcripts. This initial round of coding generated a wide range of preliminary codes, such as \textit{“Lesson Planning”} (use of genAI to generate instructional materials), and \textit{“District Policy”} (references to genAI-related policies or administrative decisions). The team then met to review and compare codes, discussing representative excerpts and resolving differences through collaborative refinement. For example, codes such as \textit{“Efficiency”} (mentions of genAI increasing productivity) and \textit{“General Advantages”} (broad references to genAI’s helpfulness) were refined into the more specific code \textit{“Time Saving”}, capturing mentions of genAI saving time for tasks like lesson planning or grading. These collaborative discussions resulted in a preliminary codebook. The team then independently applied the codebook to two additional transcripts, refining it further during subsequent discussions. This process was repeated for a total of four rounds: the revised codebook was applied to two new transcripts in each round, with the resulting discussions guiding further refinements, until the final version of the codebook was agreed upon by the coding team.

The final codebook consisted of 10 top-level codes and 45 subcodes. Consistent with TRM, the codes captured three interrelated dimensions of teachers’ genAI integration. First, they reflected teachers’ value-driven orientations, including their attitudes toward genAI (e.g., \textit{Critical}, \textit{Conditional}, \textit{Embracing}), perceived benefits (e.g., \textit{Time Saving}, \textit{Burnout Prevention}, \textit{Extra Support for Students}), and concerns (e.g., \textit{Plagiarism}, \textit{Over-Reliance}, \textit{Loss of Human Connections}). Second, the codebook represented the contextual nature of teachers’ decision-making, capturing how institutional policies, classroom dynamics, and individual backgrounds shaped choices about when and how to use genAI (e.g., \textit{Decision-Making Practices}, \textit{District Policy}, \textit{Teacher Background}). Finally, the codes highlighted teachers’ perceptions of possibility, as reflected in codes such as \textit{Future AI Use}. With the final codebook established, the team systematically applied it to the full dataset. We organized the resulting codes into overarching themes through multiple iterative rounds of refinement and discussion, drawing on our research questions and theoretical framing. After finalizing the themes, the first author revisited the entire dataset to select representative quotes that best illustrated each theme.

\begin{table*}[ht]
\centering
\caption{Overview of self-reported teacher information, including current role, years of professional experience, race/ethnicity, gender, grade-level setting, genAI use frequency, and tools used.}
\renewcommand{\arraystretch}{1.5}
\setlength{\tabcolsep}{5pt}
\rowcolors{2}{gray!15}{white}
\normalfont
\resizebox{\textwidth}{!}{%
\begin{tabular}{@{}p{1.2cm}|p{4.2cm}|p{1.3cm}|p{1.8cm}|p{1.2cm}|p{1.4cm}|p{1.2cm}|p{1.5cm}|p{1.5cm}|p{3.5cm}@{}}
\toprule
\textbf{ID} & \textbf{Current Role} & \textbf{Years of Teaching} & \textbf{Years in Current Role} & \textbf{Grades} & \textbf{Age} & \textbf{Gender} & \textbf{Race} & \textbf{AI Use Frequency} & \textbf{AI Tools Used} \\
\midrule
P01 & Classroom teacher & 11+ & 11+ & K–5 & 35–44 & Woman & Black, Native American & NA & Canva, ChatGPT, Magic School AI, Adobe AI, Gemini \\
P02 & Classroom teacher & 1–2 & 1–2 & K–5 & 25–34 & Woman & Hispanic or Latino & NA  & ChatGPT, Brisk Learning, TalkingPoints \\
P03 & Special Education, Staffing Chair, 504 Coordinator & 11+ & 3–5 & K–5 & 35–44 & Woman & White & NA & Magic School AI, ChatGPT, Canva \\
P04 & STEM teacher & 6–10 & 3–5 & K–5, 6–8 & 35–44 & Man & Asian, White & NA & Magic School AI, ChatGPT, Gemini \\
P05 & STEM teacher & 11+ & 1–2 & 9–12 & 55–64 & Man & White & Few times a Week & ChatGPT, Gemini, Copilot \\
P06 & CTE Media teacher & 11+ & 11+ & 9–12 & 35–44 & Man & Prefer not to respond & Daily & ChatGPT, Adobe AI \\
P07 & Science/Social Studies teacher & 3–5 & 1–2 & 6–8 & 25–34 & Man & White & Few times a Week & ChatGPT, Gemini \\
P08 & Math teacher & 11+ & 3–5 & 6–8 & 45–54 & Woman & White & Daily & ChatGPT, TalkingPoints \\
P09 & Classroom teacher & 3–5 & 3–5 & K–5 & 18–24 & Woman & White & Few times a Week & Magic School AI, ChatGPT, TalkingPoints \\
P10 & Science teacher & 1–2 & 1–2 & 9–12 & 18–24 & Non-binary & White &  Once/ Week & Magic School AI, ChatGPT \\
P11 & CLDE Teacher Leader & 11+ & 1–2 & 9–12 & 45–54 & Woman & Native American, White & Daily & EdPuzzle, Magic School AI, EduAid, ChatGPT, Diffit, Gemini, Moat, Screencastify, Pear Deck, Claude, Read Along, Kahoot \\
P12 & CLDE Teacher Leader & 11+ & 3–5 & 6–8 & 25–34 & Woman & Hispanic or Latino & Few times a Week & Magic School AI, TalkingPoints \\
P13 & Social Studies teacher & 11+ & 11+ & 6–8 & 35–44 & Woman & White & Once/ Month & Magic School AI, ChatGPT, Diffit \\
P14 & Substitute Teacher, Case Manager & 11+ & 1–2 & K–5, 9–12 & 45–54 & Man & White & Daily & Magic School AI, ChatGPT, Claude \\
P15 & Classroom teacher & 1–2 & 1–2 & 9–12 & 18–24 & Woman & White & Few times a Week & ChatGPT \\
P16 & Social Studies teacher & 3–5 & 1–2 & 9–12 & 25–34 & Non-binary & Black, White & Few times a Week & ChatGPT, Brisk Learning, Gemini, HeyGen \\
P17 & Science teacher & 11+ & 6–10 & 6–8 & 35–44 & Woman & White & Few times a Week & ChatGPT, Magic School AI, Pear Deck \\
P18 & Math teacher & 3–5 & 3–5 & K–5 & 25–34 & Woman & Black, White & NA & NA \\
P19 & Special Education teacher & 11+ & 11+ & K–5 & 35–44 & Woman & White & Daily & ChatGPT, Canva \\
P20 & CLDE Teacher Leader & 11+ & 6–10 & 9–12 & 45–54 & Woman & White &  Once/ Week & Magic School AI \\
P21 & Literacy teacher & 11+ & 1–2 & 6–8, 9–12 & 45–54 & Woman & White &  Once/ Month & Magic School AI \\
P22 & Classroom teacher & 11+ & 3–5 & K–5 & 35–44 & Woman & White & Rarely & Magic School AI, ChatGPT \\
\bottomrule
\end{tabular}%
}
\label{tab:teacher_participant_roles}
\noindent\parbox{\textwidth}{\small
\textit{Note.} Grades are labeled as follows: K–5 (elementary), 6–8 (middle), 9–12 (high school).
}
\end{table*}

\section{Findings}
To address our research questions about how teachers perceive the introduction of genAI tools (RQ1) and what opportunities and challenges arise as schools introduce these tools (RQ2), we organize our findings around four key areas. First, we examine the drivers that motivate teachers to adopt genAI, revealing both the opportunities they perceive and the social dynamics that shape uptake. Second, we explore the sources of reluctance and resistance that highlight key challenges in implementation. Third, we analyze how teachers actively negotiate boundaries in their genAI use, demonstrating the complex perceptions and professional considerations that guide their decision-making. Finally, we examine the conditions teachers believe are necessary for sustainable integration, which speaks to both the ongoing challenges and the systemic opportunities they envision. Across these areas, we highlight how teachers made situated decisions shaped by institutional, social, and ethical constraints. To preserve anonymity, participants are identified by a “P” followed by a unique ID (e.g., P06, P14). We also provide general teaching background, including the grade band they primarily teach (elementary, middle, or high school), their subject area, and their total years of teaching experience. For example, “P17 (Middle School Science, 11+ years of experience)” refers to a teacher who has taught middle school science for more than a decade.

\subsection{Drivers of Generative AI Adoption}

\subsubsection{Institutional Incentives and Policies}
Participants emphasized the role of district policies and communications in legitimizing and motivating their adoption of genAI tools. Formal signals from the district, whether through explicit encouragement, professional development offerings, or the assurance that use would not lead to reprimand, played a critical role in shaping teachers’ willingness to experiment. For example, P17 (Middle School Science, 11+ years of experience) described how district endorsement shifted her stance from initial hesitation to greater openness and continued use of genAI tools,\textit{“This year the district did the push for it. So then I was like, oh, good. They want us using it, so I won't get in trouble for using it, so I'll keep using it."} Teachers also described how institutional incentives, such as salary advancement courses that provide credits toward pay increases, encouraged them to try genAI during professional development. P12 (Middle School CLDE, 11+ years of experience) shared that these incentives often tipped the balance when time was scarce, \textit{“There was like a salary advancement course. So when there’s like an incentive tied to it that makes it more like, okay, I’m open to trying it out. And then if you can see what it actually looks like successfully, then I feel like it’s easier because I think sometimes at first it’s kind of like I don’t have time for this.”}

\subsubsection{Peer Sharing and Validation} Teachers described the role of peer networks in normalizing and encouraging genAI use. Rather than adopting genAI in isolation, they often recounted moments of discovery and exchange with colleagues, where trying out genAI became a shared endeavor. For example, P04 (Elementary and Middle School STEM, 6-10 years of experience) traced a series of encounters that moved him from casual exposure to becoming an advocate for genAI in his school. \begin{quote}I was introduced by our social studies teacher — ‘Hey, have you seen this? Wonderful.’ And then around that time, our district tech person who I am friends with pointed out that BPS has a subscription. And over summer, I was kind of playing around with it, and it just like blew my mind. I was all in at that point. And I showed another teacher this and I gave her kind of a quick ten second elevator pitch on it. And then she's like, you should run a PD. So I ended up running a PD for our school.\end{quote} 

Peer networks also extended beyond the walls of the school. P09 (Elementary School Classroom Teacher, 3-5 years of experience) explained how online communities helped spark her own experimentation, “\textit{I hear a lot about stuff through like TikTok or other social media and I honestly just saw a lot of other teachers being like, ‘Hey, look at how cool this tool is.’ And I kind of started incorporating it into some of my practices.}” Hearing from trusted colleagues helped counter the perceived stigma that genAI use was thoughtless or lazy. As P13 (Middle School Social Studies, 11+ years of experience) put it, “\textit{When [other teachers] talk to those of us who have used it successfully...they think we’re just plugging in and using no thought. It’s like, no, I’m skipping the steps that are just bureaucratic nonsense [to] save time.}” Lastly, sometimes, peer validation came from unexpected places, including administrators who had previously discouraged genAI. For example, P03 (Elementary Special Education, 11+ years of experience) recalled, “\textit{My boss told me I couldn’t use it last year. I sat down and he's like, I don't know how to respond to this stupid email. And I said, put it in ChatGPT and say, make it nice. And he did it and he was like, what? This is so perfect. And I'm just like, you're welcome.}”

\subsubsection{Burnout Prevention and Sustainability} Participants described the relentless pace of teaching, long hours, heavy paperwork, and the cognitive demands of responding to students and families. Thus, for many teachers, the motivation to try genAI was about protecting their ability to show up fully for students without being consumed by their work. P12 (Middle School CLDE, 11+ years of experience) captured this perspective by linking genAI adoption to burnout prevention, “I don't have to take so much time doing brainstorming for lessons by myself. I can use technology to help me...I wanna save time because I don't wanna burn myself out.” Similarly, P13 (Middle School Social Studies, 11+ years of experience) described the overwhelming scope of their workload and framed genAI as a way to sustain good teaching, \begin{quote} Our jobs are like 80 hour week jobs...the amount of time grading and also like creating things, the writing lesson plans, it's so much that is not the best use of your time. So then you can't be the best teacher you can be when you're in the classroom with kids. And it's like, well, what if it didn’t have to be?...It’s kind of like having an assistant or someone who could deal with some of the more bureaucratic stuff. \end{quote} For others, the motivation came from wanting relief from the emotional strain and mental fatigue that made teaching unsustainable. P11 (High School CLDE, 11+ years of experience) described turning to genAI as a way to reduce stress and lighten the cognitive load that often led to burnout, \textit{“The most significant benefit that AI has brought to my life as a teacher is having like, work-life balance. It has decreased my stress like 80 fold because I am able to have a thought partner. Teachers are really isolated, even though we work with people constantly...When I'm exhausted, it gives me support and help with ideas. The most important thing for me is the decrease in stress, decrease in cognitive load.”}

\subsubsection{Broadening Instructional Approaches} Several teachers described adopting genAI because it gave them a sense of creative possibility in their own practice. They spoke about using it as a playground for testing ideas, experimenting with new modalities, and re-imagining instructional approaches that felt fresh and exciting. For example, P01 (Elementary Classroom Teacher, 11+ years of experience) explained, \textit{“[My] students are researching Colorado historical figures. So what's cool in Magic School is...they can have a conversation with the historical figure through the chatbot! So if they're studying like Dr. Martin Luther King...they can have a conversation like, when were you born?''} For other teachers like P06 (High School Media Teacher, 11+ years of experience), the appeal was in extending his own instructional toolkit, \textit{“I ask ChatGPT, what’s new here? What can I use AI to support this lesson with? That does help me fill in the gap sometimes, especially when I’ve taken on new classes and had to learn as quick as I could. It’s opened my eyes to how many things could help us.”} This spirit of experimentation also reached beyond day‑to‑day classroom lessons, shaping how some teachers designed and delivered instruction more broadly. For example, P16 (High School Social Studies, 3-5 years of experience) noted, \begin{quote} I'm a professional development content creator for my school district as well. So I create micro PDs that are roughly like seven to 10 minutes long. And I use HeyGen to do that because I don't like necessarily being on camera all the time. So I write my script out and I choose an avatar from Heygen and the avatar then presents my professional development. Next year, I’m gonna be using it even more, so like when I’m absent, I’m gonna have Heygen do my lessons for me so that my substitute teacher can just push the play button on the video.\end{quote}

\subsection{Reluctance Toward Generative AI Use}

\subsubsection{Reluctance Due to Time Barriers}  
Many teachers described their lack of time as a central reason for resisting deeper engagement with genAI. Even when they acknowledged its potential benefits, they emphasized that full schedules and the effort required to learn new routines made experimentation feel burdensome. For example, P09 (Elementary School Classroom Teacher, 3-5 years of experience) explained, \textit{“It's just like lack of time. Like we don't really get much planning time and it would be a new tool to learn, so we would have to take the time personally to learn how to use it and where to find everything.”} The time consuming nature of prompting also contributed to frustration. As P02 (Elementary School Classroom Teacher, 1-2 years of experience) noted, \textit{“If I put in a prompt and it doesn't do it, then I'll have to like sit there and like edit the prompt over and over and over again...I just move on.”} Similarly, others connected these challenges directly to a sense that the effort outweighed the payoff. For example, P16 (High School Social Studies, 3-5 years of experience) described how assessment design with genAI often fell short, \textit{“When I'm writing test questions and assessments using Gemini...for the most part no AI can read images. So when I'm trying to write a test and I have like a graph, I have to describe the graph, and typically it doesn’t know what my description is exactly saying. It doesn’t work for me at all...I then put a basic description down and have it write the question, then I edit the question heavily to make it actually fit what the chart is truly saying.”} 

\subsubsection{Reluctance Due to the Stigma of “Cheating”}
For some teachers, reluctance to use genAI stemmed from the social and professional stigma attached to its use. Rather than being seen as a helpful resource, genAI was often perceived by students, parents, and colleagues as a form of “cheating” or taking the easy way out. Teachers described moments when this stigma surfaced in classroom interactions. P14 (Elementary and High School Substitute Teacher, 11+ years of experience) recalled, \textit{“Every once in a while a student will come and see that I'm using ChatGPT, and they're like, ‘You're cheating’”} Similarly, P10 (High School Science Teacher, 1-2 years of experience) recounted how students detected genAI involvement in their materials, \textit{“I did one time give my students some AI video questions and I think one of them looked at that and went, ‘Did you use ChatGPT to write this?’...and I was like, yeah. And she’s like, ‘That makes sense. It’s weirdly worded.’”} Such encounters reinforced teachers’ worries about being seen as cutting corners. Parent attitudes added another layer of pressure. As P02 (Elementary School Classroom Teacher, 1-2 years of experience) explained, \textit{“We do have a couple of difficult parents that...if they knew we were using it, they would be very upset and just think that we're not supporting their kids the way they need to be supported.”} For some teachers, these perceptions provoked deeper questions about the value of their expertise and their professional identity. As P20 (High School CLDE, 11+ years of experience) reflected, \textit{“I think there’s a certain aspect of it that feels like it’s kind of cheating, right? Like you’re in education, you should be able to...think and come up with ideas on your own...If AI’s doing all this work, then what am I doing?”}

\subsubsection{Reluctance Rooted in Preference for Tradition}  
For several teachers, established practices offered a sense of reliability and continuity, making experimentation with new tools feel undesirable. As P12 (Middle School CLDE, 11+ years of experience) explained, \textit{“I think it's like anyone, right? You don’t really like change. So sometimes it’s like, why would I do something [new]...what I’ve been doing has been working, or it seems like it’s been working, or like, [you] kind of already have the material so [you] just kind of reuse it and adjust.”} Teachers also observed that some of their colleagues avoided genAI because of its unfamiliarity, preferring the certainty of existing workflows. As P20 (High School CLDE, 11+ years of experience) noted, \textit{“We have people who are genuinely concerned and afraid of it and don’t wanna adopt it, and don’t want anything to do with it. They have all their plans in place and...there’s no reason to do anything differently.”} Some framed this preference in explicitly “old school” terms, emphasizing low-tech, hands-on teaching methods over digital experimentation. As P03 (Elementary Special Education, 11+ years of experience) shared, \textit{“From my teaching perspective, like I’m not really using it actively...I’m pretty low tech, like old school in terms of teaching. Like, kids have whiteboards in my room, they have a dry erase marker. We write words, we read books. I have tried AI to create decodable texts, but I have not found it to be very good at that yet.” } 

\subsubsection{Reluctance Due to Environmental Concerns}
For a small subset of teachers, reluctance to embrace genAI was rooted in concerns about its environmental impact and the broader ethical costs of widespread adoption. Several described scaling back their own use after learning about the resource intensity of these systems. As P09 (Elementary School Classroom Teacher, 3-5 years of experience) explained, \textit{“I definitely used to use it more and then I saw some stuff about how much water is used in AI and so I pulled back my use on it.”} Others voiced broader ethical unease about AI’s contribution to climate change. P13 (Middle School Social Studies, 11+ years of experience) reflected, \textit{“My biggest concern is what I've seen about the carbon footprint of it and the environmental impact of it. We're already destroying the planet at a rapid rate, so like are we just accelerating that?”} Some positioned themselves as openly critical of genAI because of these issues. As P10 (High School Science Teacher, 1-2 years of experience) shared, \textit{“I might be the most vocally against it in like my school because I'm aware of like the environmental horrors. I’ve told my coworkers about how much water is being used to like cool down the servers, but I haven't heard of anyone like trying to actively change administration or district minds about it.''}

\subsection{Navigating Boundaries in AI Use}
Beyond clear cases of adoption or reluctance, many teachers described actively negotiating boundaries around their use of genAI. They framed these boundaries as necessary to preserve their professional judgment, avoid perceptions of “cheating,” and ensure that students remained engaged in meaningful learning.

\subsubsection{Preserving Professional Voice and Judgment}
Many teachers drew boundaries around genAI use to ensure that their professional voice and judgment remained central to their work. While they welcomed genAI as a source of support, they worried that over‑reliance on AI‑generated content could make their teaching feel impersonal, generic, or detached from students’ needs. Several teachers emphasized that using genAI without critical review was unacceptable. As P07 (Middle School Science/Social Studies, 3-5 years of experience) explained, \begin{quote}If you're just saying, `Make me a lesson plan on this,' and you don't go through it, and you just do it -- that's not okay. It's help, it's support, it's taking some of the load off our back. It's to make us better educators not just rely on it. If you're too reliant and you're just spitting it out there, it's gonna hurt our kids.\end{quote} Teachers also highlighted the importance of reviewing and adapting outputs to reflect their own judgment. As P16 (High School Social Studies, 3-5 years of experience) put it, \textit{``It is a great way to start writing a test or giving feedback to students, but you need to go back in and edit, make sure that it's actually aligned to what you want the student to work on, and then grade the writing yourself. That's my framing: it’s a starting point.''} Others described feeling uneasy when genAI outputs sounded generic, inauthentic, or inconsistent with their own communication style. For example, P20 (High School CLDE, 11+ years of experience) recounted noticing how her principal’s tone shifted after incorporating genAI into school communications, \textit{“I will never forget the moment where I was like, he's using AI because you could tell that his communication had changed."} In response to similar concerns about sounding inauthentic, P12 (Middle School CLDE, 11+ years of experience) shared, \textit{``I [take] whatever it generates and then just review it, because  at the end of the day, you still want it to sound like you. And you still want it to be approachable.'' }

\subsubsection{Negotiating Trust in AI Collaboration}  
Many teachers described a process of careful decision-making shaped by concerns about the reliability of genAI. Several pointed to factual errors as a key reason for caution, particularly in subjects that demand precision. P13 (Middle School Social Studies, 11+ years of experience) reflected, \textit{“We’ve all seen and heard ways that ChatGPT can be wrong about, especially in social studies. Like lots of wrong dates, wrong, you know what I mean?... It is not necessarily a trusted source, it’s just a majority source.”} Others emphasized the need for vigilant fact-checking to guard against misinformation. P21 explained, \textit{“Even AI can get fooled with fake news... for me nowadays, it’s almost a hundred percent fact check before I post anything.”} Similarly, teachers extended their caution to worries about bias and the potential for genAI to spread harmful or politically skewed information. As P17 (Middle School Science, 11+ years of experience) noted, \textit{“Just thinking of our political climate, like if everything that it spewed out was just like wrong and not real science... that scares the crap outta me.”}  P17 further reflected that these concerns were not abstract, noting that students already bring misinformation into the classroom. They explained, \textit{“You’re already dealing with that in the classroom sometimes. Like, oh, my family says this. And it’s like, okay, hey, this is science. We deal with scientific facts only.''} This according to P17 made it especially important to teach critical habits of verification, \textit{“You can’t blindly trust AI. So instilling that into kids early on, ’cause as a 44-year-old new to it, I was like, oh you can’t, oh okay duh. I should know that, but I didn’t.”}  

\subsubsection{Protecting Student Data Privacy}
Teachers also described drawing boundaries around the use of genAI due to concerns about privacy and data security. Even when they saw potential benefits, many hesitated to input sensitive student information or use platforms they felt were not transparent about data practices. For example, P07 (Middle School Science/Social Studies, 3-5 years of experience) described being meticulous about redacting personal details before pasting student work into genAI tools, 
\begin{quote}There have been a few times where I may be copying, like student work over into AI for it to help me grade. And in those situations I have to be very careful that I'm purposely only pasting the content of the assignment and not a name or personal information or even drawings the kid did...Because of that I can't just like scan the document and upload it. Sometimes I would rather just grade myself, because usually grading is the only time that I have to pay attention to student information or student data. Other than that, I try to avoid talking about anything confidential and just leave it strictly to curriculum when I'm talking to AI.\end{quote} For teachers working in sensitive contexts, the stakes felt even higher. P11 (High School CLDE, 11+ years of experience)  explained how they leveraged genAI to support students' Individualized Education Program (IEP), which provides accommodations for students with special needs, while avoiding the use of identifiable data. \textit{``I attend IEP meetings very frequently, and we provide test data in reports we create. I don’t put in personally identifying information, but I’ll enter the scores into Gemini and ask, `What are some strengths of these scores? What are are some areas to work on for these scores?'} It doesn't have the kid's name or anything attached to it, but it will tell me like, oh, the reading shows this trend.'' At the institutional level, concerns about compliance with student privacy laws further shaped teachers’ decisions. P04 (Elementary and Middle School STEM, 6-10 years of experience) noted how district protocols limited which tools could be used, \textit{``A lot of [admins] are like, what are the privacy concerns? BPS is really starting to crack down — if a website doesn’t have the insurance or the protocol set in place, we’re not allowed to use them for sign‑in with Google. I know there's a lot of admin that are really concerned about [this], so they're like, let’s stick to the curriculum we’re already using.''}

\subsubsection{Boundaries for Student Exposure and Responsible Use}
Teachers noted the need to be intentional about how, if at all, they introduced genAI into student learning. This negotiation was shaped by the dual goals of empowering students to explore new tools while preventing over‑reliance or misuse. As P07 (Middle School Science/Social Studies, 3-5 years of experience) explained, \begin{quote}Usually I start with, what am I measuring? If I'm measuring their ultimate learning on one skill, I'm not gonna allow it. But if I'm measuring growth, AI is okay. As long as we are working on growth, AI can help you grow. But once we get to mastery time, you need to be able to show me that you can do it without that support. That way I know they're not relying on it for like the final results. Those are my two big separators.\end{quote} For teachers working in special education, boundaries often centered on balancing support with skill development. P19 reflected (Elementary Special Education, 11+ years of experience), ``\textit{They use speech‑to‑text already, so why not take it to the next level and say, `This is the idea I want to get across. How can I word this?' I don’t think it’s not teaching them; I think it’s teaching them a tool that will be widely available as they get older. They do need to learn conventions of English, but you can do both.}'' Others emphasized teaching students to think critically about their use of genAI. For example, P03 (Elementary Special Education, 11+ years of experience) described how they coached their fourth‑graders to reflect on authorship,
\begin{quote}
I tell my students, I don't care if you use it. What I wanna know is that you thought about what it was producing and why. You can’t just say, “Write me this response to this text,” because that’s not your ideas. I care what your ideas are. Now, if you need help with how to say it, you can say, “Help me say this in a better way.” Cool, I love that. I’m super explicit with my fourth graders on how to use AI, because a lot of my kids can’t write, they can’t read. And they’d be using AI anyway.
\end{quote}

\subsection{Conditions for Sustainable AI Integration}
In reflecting on their experiences with genAI, many teachers emphasized that their individuals efforts could only go so far without systemic supports. They highlighted gaps in institutional policies and guidance, professional development opportunities, and clear expectations for students that shaped how and how well genAI could be used in their district. At the same time, they voiced concerns about what might be lost if integration proceeded without attention to the human and relational dimensions of teaching. Together, these reflections outlined conditions they saw as essential for the responsible and sustainable use of genAI in schools.

\subsubsection{Clear District Policies and Guidance}
Teachers repeatedly emphasized that effective genAI integration depended on clear, consistent guidance from their districts. Many felt uncertain about the rules that governed their use of genAI, which left them wary of experimenting too freely. As P22 (Elementary Classroom Teacher, 11+ years of experience) explained, \begin{quote}I definitely think knowledge of how it can be used for each district would be nice as far as what the district rules are and how teachers can implement it. Because I know there are different districts, different states that kind of have their own take on it...Just knowing those rules and regulations would definitely be helpful and everybody would feel a little bit safer in using those things.\end{quote} In the absence of such clarity, some teachers described taking their own precautions to remain professional and protected. As P03 (Elementary Special Education, 11+ years of experience) put it, \textit{“I’m using my school account for a reason. Like if the district ever wanted to say, ‘I need to look at what you’ve been doing,’ I could literally pull up my school account and say, ‘Here’s everything I’ve ever done.’ I’m pretty intentional about that. I think people need to know how to protect themselves to be professional.”}  For others, the lack of guidance raised questions about ethics, particularly around plagiarism and authenticity. P19 reflected (Elementary Special Education, 11+ years of experience) reflected, 
\begin{quote}
Teachers [sometimes think] that [AI use] is some sort of plagiarism...I think getting clear on what is ethical, what is allowed and what isn’t, and really defining: Is AI plagiarism or not? Where are those boundaries? Where are those lines? Because there really is not a lot of clarity in terms of AI. It’s more of just a `go explore it and use this in a way that might help you,' which is great for some people and not so great for others who need more boundaries and guidance.
\end{quote} This need for clarity extended to the approval and vetting of tools. For example, P17 (Middle School Science, 11+ years of experience) suggested, \textit{“They started off giving us a program [Magic School]. So I think they should have a list of approved ones they want us using for stuff in the classroom.”} Similarly, P21 (Middle and High School Literacy, 11+ years of experience) described the confusion that arose when tools appeared without sufficient communication, noting, \textit{“When Magic School AI first came out, I wasn’t sure it was legitimate. I wasn't sure if it was something that was fake or, some company trying to sell me something. Being a little more open as to what they’re buying and what are the benefits of it would help.”} 

\subsubsection{The Need for Professional Development}
Participants described the current landscape as confusing and uneven, with little guidance on how to use genAI in ways that were pedagogically sound. Several emphasized the importance of structured training led by educators who understood classroom realities. As P04 (Elementary and Middle School STEM, 6-10 years of experience) explained, \textit{“They should have people that are teachers that understand the teaching process and how to integrate it. That way there’s more buy‑in, because I think there’s still that fear of the unknown with some of our more veteran teachers.”} Others stressed the need for clear and accessible entry points to reduce hesitation. P07 (Middle School Science/Social Studies, 3-5 years of experience) suggested, \textit{“A good one‑pager detailing when AI is useful, how it can help teachers, and when it might be too much or not appropriate for students...just to get educators started and allow them that familiarity, so it’s not such a scary thing.”} Similarly, P19 reflected (Elementary Special Education, 11+ years of experience) pointed to both practical and ethical dimensions of training, \textit{“Something I think would help is examples of prompts for different purposes, and I think ethics training on using AI, it’s just so new that people don’t know what is right and wrong.”} Teachers also emphasized that without targeted professional development, many colleagues would continue to hold back from using genAI. P16 (High School Social Studies, 3-5 years of experience) described how limited understanding fueled fear and resistance, underscoring why training felt urgent,
\begin{quote}
The number one reason people don’t want to use AI is because they don’t understand it. Districts and schools need to provide accessible and detailed training about what genAI actually is, what it does, and how it works. A lot of people are just worried, “Oh my gosh, it’s going to be like the Terminator and it’s going to be so bad.” But we’re nowhere near that level of AI. People don’t understand the nuances that many different types of AI exist out there. So that's the number one thing that districts and schools need to do.
\end{quote}

\subsubsection{Building Critical AI Competencies in Students}
Teachers highlighted that preparing students to navigate a world with genAI is an essential condition for sustainable integration. They argued that school districts making decisions top down and restricting access or avoiding the topic would leave students unprepared for the tools that will shape their futures. As P20 (High School CLDE, 11+ years of experience) put it, ``\textit{I don’t think we should just put a hard, fast rule like absolutely no AI, we’re not using it, we’re blocking anything. AI is the wave of the future. If we say we’re gonna make them college and career ready, we need to teach them the tools, but we need to teach them the ethics behind it as well, so they can make good decisions.}'' Several teachers noted that the absence of structured guidance had already led to concerning patterns of use. For example, P07 (Middle School Science/Social Studies, 3-5 years of experience) described one student whose dependence on genAI had become disruptive, 
\begin{quote}
I have one student who is probably single handedly keeping ChatGPT in business, because she is glued to it unhealthily. She’s not using it for school, she’s just creating a story every day, all day. And if we block her, she throws an absolute tantrum.
\end{quote}
To counter these risks, participants stressed that genAI literacy should not be treated as optional or left to chance. They argued for explicit lessons that move beyond technical use to include reflection on ethics and authorship. For example, P08 noted, \textit{“We should really be encouraging kids to challenge what they read, what they listen to, what they hear.”} In the absence of district‑wide frameworks, some teachers created their own entry points for genAI literacy. P04 (Elementary and Middle School STEM, 6-10 years of experience) recounted,
\begin{quote}
I did a little mini lesson on AI and found a website that you have to click on, which one is artwork is created by human and which is not. And then we just shared out scores, and then it led into a conversation of who owns the artwork. It really kind of opened their eyes because prior to that, they're like, let me do this with [AI]. Let me have it answer my questions for me. It's kind of cool when you get a middle schooler to think like a little more outside of their own personal bubble.
\end{quote}
These teacher‑led efforts underscored a broader need that without structured curricula and district guidance, students’ exposure to genAI literacy remains inconsistent and dependent on individual teacher initiative.

\subsubsection{Preserving the Relational Dimensions of Learning}
Teachers argued that for genAI adoption to be sustainable, it must not erode the human relationships that make learning meaningful. They stressed that authentic connection, trust, and opportunities for social learning remain central to education, even as digital tools expand. As P13 (Middle School Social Studies, 11+ years of experience) explained, 
\begin{quote}
A machine can give you information, but most students we know are not able to get information from something that’s just printed out for them and put it into their heads. You need a relationship. Some kids can do online school or read a book and teach themselves, but that's like 2\%. Most kids need a social environment to do it.
\end{quote}
Participants cautioned that if schools leaned too heavily on genAI, they risked undermining the very qualities that sustain students’ growth. P06 (High School Media Teacher, 11+ years of experience) reflected, \textit{“Worst case is schools commit to almost embracing it too much, and it takes away what makes school special -- the connections and people. If we move more online, we lose that personal aspect of learning.”} Others noted that the challenge was not whether to use genAI, but how to balance it with intentional opportunities for human connection. As P03 (Elementary Special Education, 11+ years of experience) put it, \textit{“One of the major negative things I’ve seen with increasing technology is that kids really crave personal connection. Since COVID, we’ve realized the necessity of human connection and how important it is to use technology intentionally to create it.”} Together, these accounts suggest that for genAI to be integrated sustainably, it must serve as a tool that supports, rather than replaces, the relational and social dimensions of learning. 

\section{Discussion}

\subsection{Navigating Tensions in Generative AI Integration}
Prior work on educational technology integration highlights that teachers' adoption processes are rarely linear and are shaped by ongoing negotiations among their pedagogical values, institutional expectations, and classroom realities \citep{reich2020failure, rafalow2020digital, ames2015charismatic}. Guided by TRM \citep{kopcha2020process} and a sociotechnical lens, our study examined how teachers made sense of and integrated genAI in their classroom practice. Our findings reveal that genAI amplifies these negotiation processes by introducing interpretive work that requires teachers to continuously assess system capabilities, evaluate contextual appropriateness, and reconcile genAI outputs with their instructional objectives. Through our analysis, we identify two interconnected tensions that fundamentally shape teachers' genAI integration processes. 

\subsubsection{Navigating the Tension between Relief and Displacement}
Teachers approached genAI with curiosity and a sense of possibility, often drawn to its potential to reduce time-intensive tasks and expand their instructional repertoire. At the same time, teachers emphasized that using genAI effectively involved crafting precise prompts, reviewing outputs critically, and adapting results to meet student's diverse needs \citep{Elsayary2025ExaminingTR, Biton2025LearningTC, Novita2025AIIL}. In this way, experimentation with genAI became a site where teachers negotiated the tension between relief and displacement. While genAI offered efficiency, teachers enacted practices that reaffirmed their expertise, ensuring that automation supported rather than supplanted their intellectual and professional authority. Nolan et al. \citep{GreeneNolanPhD2024TeachingPG} characterize this kind of interpretive labor \citep{StaudtWillet2024EducatorsIL, Ankitdhamija2024UnderstandingTP} as a form of professional expertise, highlighting how teachers’ careful oversight is essential for the responsible use of genAI in classrooms. Likewise, Xie et al. \citep{Xie2024CodesigningAE} and Mah et al. \citep{Mah2024BeyondCE} show how teachers rely on flexible, context-sensitive heuristics, rather than fixed protocols, to integrate emerging tools.

Teachers' evaluative work also extended to protecting the relational core of learning. They valued genAI when it created space for their relationships with students to flourish \citep{Ferman2021ArtificialIT, Lawrie2023EstablishingAD}. At the same time, teachers worried about losing opportunities for authentic connection with students \citep{Cipolla2024GenerativeAI}. These concerns echo longstanding research on the centrality of relational dynamics in teaching \citep{Windschitl2002TracingTU, Gu2014TheRO}. Simon and Johnson \citep{simon2015teacher}, for example, demonstrate that the quality of teachers’ social interactions with students and peers is the strongest predictor of retention in high-poverty schools. Similarly, Ball \citep{ball2022reimagining} argues that the power of teaching lies in its deeply social nature, where meaning is co-constructed rather than delivered. TRM \citep{kopcha2020process} helps make sense of these negotiations, highlighting how teachers assess both possibilities and constraints of new technologies in light of their pedagogical commitments. Additionally, our findings also extend prior scholarship on how technologies reconfigure, rather than replace, teachers' professional practices \citep{Varanasi2019HowTI, Kozma2003TechnologyAC}.

\subsubsection{Navigating the Tension between Mediating Student Use and Building Teacher Expertise} 
A second tension emerging from our findings concerns teachers’ dual roles as mediators of genAI for their students while also navigating these systems as novice users. As with earlier technologies such as MOOCs \citep{wang2015content, baturay2015overview}, or intelligent tutoring systems \citep{graesser2012intelligent, anderson1985intelligent}, genAI entered classrooms faster than professional development or policy infrastructures could respond, leaving many educators unprepared \citep{Kohnke2023ExploringGA, Brando2024TeacherPD, Prather2024BeyondTH}. In the absence of structured guidance, some teachers in our study created their own activities to help students interrogate genAI’s limitations, authorship, and biases, reflecting calls in HCI and education research to integrate genAI literacy into classrooms \citep{VanBrummelen2020EngagingTT, Walter2024EmbracingTF}. Other participants, however, carried misconceptions about genAI or avoided it altogether. These divergent approaches were compounded by uneven support at BPS as access to professional development and resources varied widely across schools. Our findings echo prior research on educational technologies, which shows that such uneven rollouts often exacerbate existing disparities rather than resolve them \citep{toyama2015geek, reich2017good, rafalow2020digital}.

Unlike prior educational technologies like MOOCs \citep{baturay2015overview}, which are typically narrowly bounded with respect to the information delivered and the kinds of interaction available to learners, genAI places more responsibility on the shoulders of teachers to evaluate the many capabilities and forms of interaction that such models afford. Developing genAI literacy among teachers requires that teachers understand what the capabilities of a general-purpose model are; how they might be applied in an educational setting; and how such applications are likely to affect student learning long-term \citep{Walter2024EmbracingTF, Long2020WhatIA}. In cases where teachers are uncertain of how genAI will affect student learning, many will justifiably choose to avoid incorporating the technology into their classroom. Yet this may exacerbate inequalities, given the technology's potential to supplement student learning when deployed intentionally and with care \citep{Gabriel2024GenerativeAA, Valdivieso2025GenerativeAT}. Equitable access not just to genAI but to teacher literacy training is crucial for ensuring that learners receive the benefit of effective incorporation of genAI in the classroom \citep{Brando2024TeacherPD, cukurova2024ai}.

\subsection{A Sociotechnical Lens on Generative AI Integration}
The two tensions we identify — between relief and displacement, and between mediating student use and developing teacher expertise — illustrate that teachers’ work with genAI cannot be understood within the boundaries of individual classrooms alone. Teachers’ practices emerged at the intersection of multiple, interconnected layers, where individual identities, peer networks, institutional structures, and broader cultural narratives all shaped how genAI integration unfolded \citep{kaghan2001out, latour1987science, leonardi2009crossing}. Through the TRM \citep{kopcha2020process} lens, we see how these layers interact, helping us place our findings in dialogue with prior work.

\subsubsection{Individual Layer: Boundary-Setting and Relational Pedagogy}
Teachers engaged in sophisticated boundary-setting practices that preserved their professional autonomy while leveraging technological efficiency. This integration work involved developing new competencies around prompt crafting and output evaluation that reshaped rather than replaced core pedagogical skills \citep{Elsayary2025ExaminingTR, Biton2025LearningTC}. Such dynamics echo prior research showing how educators reinterpret technological innovations to fit professional norms rather than adopt them wholesale \citep{selwyn2016high}. Additionally, teachers’ emphasis on maintaining authentic relationships with students while preventing burnout suggests that integration was also motivated by protecting relational forms of pedagogy \citep{ball2022reimagining, simon2015teacher}.

\subsubsection{Social Layer: Peer Networks as Interpretive Communities}
Teachers' individual practices were deeply entangled with the social contexts in which they were situated. Peer networks and online communities played an important role in mediating experimentation, offering strategies, resources, and validation when official guidance was absent. This resonates with prior studies documenting how informal communities of practice often become critical infrastructures for educational technology adoption \citep{Forte2012GrassrootsPD, Jin2025KnowledgeA, Floris2024ArtificialII}. Yet this grassroots legitimation was fragile. Teachers also navigated skepticism from students, parents, and even other educators who questioned whether using genAI compromised authenticity or professionalism \citep{Han2024TeachersPA}. These mixed dynamics positioned peer networks not simply as information-sharing channels but as interpretive communities where meaning was negotiated and professional legitimacy was continuously managed \citep{kopcha2020process}.

\subsubsection{Institutional Layer: Fragmented Policies and Uneven Access}
District endorsement lowered barriers by signaling permission to experiment, but the absence of clear implementation guidelines left teachers to make highly individualized decisions. This unevenness echoes longstanding critiques of edtech integration, where innovations often exacerbate rather than mitigate disparities in resources and support \citep{toyama2015geek, reich2017good}. Teachers with access to professional development opportunities or strong peer networks were better positioned to explore genAI’s potential, while others avoided it altogether due to uncertainty or lack of support. This reliance on individual initiative reflects a broader structural challenge in which without coordinated frameworks, integration risks becoming uneven and unsustainable, particularly for educators working in resource-constrained contexts \citep{reich2020failure}.

\subsubsection{Sectoral Layer: Cultural Logics and Ethical Concerns}
Finally, teachers’ negotiations unfolded within broader cultural narratives about technology and innovation \citep{Curwood2014BetweenCA}. Several participants described wanting to “keep up” and experiment with new tools out of curiosity and a sense of professional growth, echoing prior accounts of teachers’ professional identities being shaped by cultural logics of innovation and technological enthusiasm \citep{laak2024generative, Wang2025ScaffoldingCI}. Yet these aspirations often collided with efficiency-first design assumptions embedded in many genAI tools, a mismatch that reflects the enduring tension between industrial logics of optimization and relational models of pedagogy \citep{selwyn2016high}. Teachers’ environmental concerns also connect local classroom practice to wider ethical debates about the sustainability of emerging technologies \citep{Henriksen2024CreativeLF, Dhiman2025EthicalCN}, extending prior critiques of digital education’s environmental footprint into the domain of genAI \citep{Shengjergji2024EnvironmentalIO}.

\subsection{Implications for Design and Policy} 

\subsubsection{Centering Teachers in Policy, Procurement, and Practice} Our findings suggest that policies around genAI integration should move beyond compliance-oriented framings (e.g., plagiarism detection) \citep{Baron2024AreAD, Lee2024CheatingIT}. Rather than treating educators as downstream implementers of pre-determined policies or tools, districts and states could create participatory governance structures in which teachers co-develop acceptable-use policies, procurement standards, and evaluative frameworks for genAI tools \citep{VanBrummelen2020EngagingTT, Xie2024CodesigningAE}. Such involvement ensures that adoption decisions better reflect classroom realities and pedagogical priorities. Additionally, requiring greater transparency from vendors, including clear documentation of data practices, bias mitigation strategies, and instructional claims, can empower teachers to make informed decisions about tool selection and classroom integration. Finally, investing in sustained, collective professional learning infrastructures, where educators collaboratively experiment with tools, interrogate risks, and share emergent practices can foster a culture of distributed expertise \citep{lu2024supporting, Brando2024TeacherPD, Jin2025KnowledgeA}. Such efforts would help to shift away from isolated, one-off trainings and toward continuous capacity-building that better aligns policy, procurement, and practice.

\subsubsection{Designing for Situated Agency} Our findings highlight the need to design genAI systems that foreground teachers’ situated expertise rather than prioritizing efficiency-driven paradigms \citep{Yin2025ResponsibleAI, Xie2024CodesigningAE}. Teachers in our study engaged in ongoing interpretive work, evaluating genAI outputs, reconciling them with classroom contexts, and adapting them to students’ diverse needs. To support this labor, genAI tools could make their underlying pedagogical assumptions explicit, allowing teachers to assess whether outputs align with their instructional goals rather than defaulting to automated recommendations \citep{Holstein2019DesigningFC}. Designing for situated agency also requires creating transparent interfaces that clarify how responses are generated, including citations, confidence signals, and potential limitations \citep{Dangol2025AIJK}. Such transparency enables teachers to critically evaluate outputs and model genAI literacy for students. Finally, genAI systems can scaffold multiple entry points that accommodate varying levels of teacher familiarity with genAI, supporting iterative learning and fostering professional agency over time.

\subsubsection{Reframing AI Literacy as Collective Capacity-Building} Prior frameworks on AI literacy have focused on developing students’ conceptual and procedural knowledge of AI—how models work, where biases emerge, and how outputs are generated \citep{payne2019ethics, solyst2023would, williams_is_2019}. More recent work in HCI and the learning sciences pushes beyond this, calling for participatory approaches that involve multiple stakeholders in interrogating AI’s assumptions and shaping its role in learning contexts \citep{Delgado2021StakeholderPI, Maas2024BeyondPA}. Building on this shift, our findings suggest that AI literacy in K–12 settings is less about acquiring discrete technical skills and more about fostering collective sensemaking distributed across teachers, students, families, and administrators \citep{druga20214as, Yang2024DesigningAA}.

To move toward this vision, collective capacity-building could take the form of distributed infrastructures that make AI literacy a shared responsibility rather than an isolated task. For example, schools could integrate collaborative digital and physical spaces, where teachers, students, and families co-analyze genAI outputs, surface hidden assumptions, and debate appropriate uses of these tools. Districts and libraries might host community-driven ``AI inquiry labs'' or design nights, enabling families and educators to learn alongside one another \citep{roque2023imagining}. Within classrooms, lightweight prompts, discussion protocols, and co-designed teaching materials could scaffold critical dialogue around genAI’s role in everyday life, connecting personal experiences with broader societal impacts \citep{lin2020zhorai, morales2024s}. Across these contexts, the goal is not to master genAI’s inner workings but to cultivate sustained, interdependent practices for understanding and shaping its use collectively.

\section{Limitations}
In focusing on the experiences of teachers, our study does not address, for the most part, the elephant in the room: that students themselves are using genAI to help them with their schoolwork. Though this must surely inform any sociotechnical view of genAI in schools, a large body of work contemporary to ours has provided insight into student experiences at various levels of instruction \citep{ammari2025students,zheng2025students,schneider2025thematic,Lee2024CheatingIT, pu2025can}, allowing us to focus attention on the relatively undertreated domain of genAI use among teachers. Moreover, we note that our study recruited K–12 teachers from a single public school district in the United States. While BPS includes a wide range of instructional settings and student populations, all participants were drawn from within a single district context. As such, the perspectives captured in this study may not reflect the experiences of educators in districts with different resource levels, governance structures, or approaches to genAI integration. All interviews were conducted remotely via Zoom. While this approach allowed for scheduling flexibility, it may have limited opportunities to observe contextual cues or develop the informal rapport often facilitated through in-person interactions. Though the findings are not intended to be generalizable, they offer timely insight into how teachers are navigating genAI in practice within a large public school system. As schools continue to adopt and regulate genAI tools, further research is needed to explore how broader institutional, cultural, and policy contexts influence adoption across a wider range of educational settings.

\section{Conclusion}

Amid the rapid change in classroom technologies induced by genAI, we offer a perspective on how teachers approach the opportunities and risks of these technologies. Our work extends beyond specific, outcome-based perspectives focused solely on student learning, instead shedding light on the intersection of genAI with the administrative, normative, and relational aspects of the profession of teaching. We believe that this holistic view can help to lay a foundation both for effective implementation of genAI in classrooms, and for policy that avoids some of the failures and inequities that have accompanied the rollout of past educational technologies.

\bibliographystyle{ACM/ACM-Reference-Format}
\bibliography{references}


\begin{thebibliography}{130}

%%% the Web address.

\ifx \showCODEN    \undefined \def \showCODEN     #1{\unskip}     \fi
\ifx \showDOI      \undefined \def \showDOI       #1{#1}\fi
\ifx \showISBNx    \undefined \def \showISBNx     #1{\unskip}     \fi
\ifx \showISBNxiii \undefined \def \showISBNxiii  #1{\unskip}     \fi
\ifx \showISSN     \undefined \def \showISSN      #1{\unskip}     \fi
\ifx \showLCCN     \undefined \def \showLCCN      #1{\unskip}     \fi
\ifx \shownote     \undefined \def \shownote      #1{#1}          \fi
\ifx \showarticletitle \undefined \def \showarticletitle #1{#1}   \fi
\ifx \showURL      \undefined \def \showURL       {\relax}        \fi
% invisible to TeX
\providecommand\bibfield[2]{#2}
\providecommand\bibinfo[2]{#2}
\providecommand\natexlab[1]{#1}
\providecommand\showeprint[2][]{arXiv:#2}

\bibitem[Acevedo(2025)]%
        {acevedo2025exploring}
\bibfield{author}{\bibinfo{person}{Katie Acevedo}.} \bibinfo{year}{2025}\natexlab{}.
\newblock \showarticletitle{Exploring the Impact of Generative AI to Mitigate Educator Burnout}.
\newblock  (\bibinfo{year}{2025}).
\newblock


\bibitem[Alibali and Nathan(2018)]%
        {Alibali2018EmbodiedCI}
\bibfield{author}{\bibinfo{person}{Martha~Wagner Alibali} {and} \bibinfo{person}{Mitchell~J. Nathan}.} \bibinfo{year}{2018}\natexlab{}.
\newblock \showarticletitle{Embodied Cognition in Learning and Teaching}.
\newblock
\urldef\tempurl%
\url{https://api.semanticscholar.org/CorpusID:149890428}
\showURL{%
\tempurl}


\bibitem[Ames(2015)]%
        {ames2015charismatic}
\bibfield{author}{\bibinfo{person}{Morgan~G Ames}.} \bibinfo{year}{2015}\natexlab{}.
\newblock \showarticletitle{Charismatic technology}. In \bibinfo{booktitle}{\emph{Proceedings of the fifth decennial Aarhus Conference on Critical Alternatives}}. \bibinfo{pages}{109--120}.
\newblock


\bibitem[Ammari et~al\mbox{.}(2025)]%
        {ammari2025students}
\bibfield{author}{\bibinfo{person}{Tawfiq Ammari}, \bibinfo{person}{Meilun Chen}, \bibinfo{person}{SM Zaman}, {and} \bibinfo{person}{Kiran Garimella}.} \bibinfo{year}{2025}\natexlab{}.
\newblock \showarticletitle{How Students (Really) Use ChatGPT: Uncovering Experiences Among Undergraduate Students}.
\newblock \bibinfo{journal}{\emph{arXiv preprint arXiv:2505.24126}} (\bibinfo{year}{2025}).
\newblock


\bibitem[Anderson et~al\mbox{.}(1985)]%
        {anderson1985intelligent}
\bibfield{author}{\bibinfo{person}{John~R Anderson}, \bibinfo{person}{C~Franklin Boyle}, {and} \bibinfo{person}{Brian~J Reiser}.} \bibinfo{year}{1985}\natexlab{}.
\newblock \showarticletitle{Intelligent tutoring systems}.
\newblock \bibinfo{journal}{\emph{Science}} \bibinfo{volume}{228}, \bibinfo{number}{4698} (\bibinfo{year}{1985}), \bibinfo{pages}{456--462}.
\newblock


\bibitem[Ankitdhamija and Dhamija(2024)]%
        {Ankitdhamija2024UnderstandingTP}
\bibfield{author}{\bibinfo{person}{Ankitdhamija} {and} \bibinfo{person}{Deepika Dhamija}.} \bibinfo{year}{2024}\natexlab{}.
\newblock \showarticletitle{Understanding Teachers' Perspectives on ChatGPT-Generated Assignments in Higher Education}.
\newblock \bibinfo{journal}{\emph{Journal of Interdisciplinary Studies in Education}} (\bibinfo{year}{2024}).
\newblock


\bibitem[Atkinson and Wilson(1968)]%
        {atkinson1968computer}
\bibfield{author}{\bibinfo{person}{Richard~C Atkinson} {and} \bibinfo{person}{HA Wilson}.} \bibinfo{year}{1968}\natexlab{}.
\newblock \showarticletitle{Computer-assisted instruction}.
\newblock \bibinfo{journal}{\emph{Science}} \bibinfo{volume}{162}, \bibinfo{number}{3849} (\bibinfo{year}{1968}), \bibinfo{pages}{73--77}.
\newblock


\bibitem[Baidoo-Anu and Ansah(2023)]%
        {baidoo2023education}
\bibfield{author}{\bibinfo{person}{David Baidoo-Anu} {and} \bibinfo{person}{Leticia~Owusu Ansah}.} \bibinfo{year}{2023}\natexlab{}.
\newblock \showarticletitle{Education in the era of generative artificial intelligence (AI): Understanding the potential benefits of ChatGPT in promoting teaching and learning}.
\newblock \bibinfo{journal}{\emph{Journal of AI}} \bibinfo{volume}{7}, \bibinfo{number}{1} (\bibinfo{year}{2023}), \bibinfo{pages}{52--62}.
\newblock


\bibitem[Ball(2022)]%
        {ball2022reimagining}
\bibfield{author}{\bibinfo{person}{Deborah~Loewenberg Ball}.} \bibinfo{year}{2022}\natexlab{}.
\newblock \showarticletitle{Reimagining American Education: Possible Futures: Coming to terms with the power of teaching}.
\newblock \bibinfo{journal}{\emph{Phi Delta Kappan}} \bibinfo{volume}{103}, \bibinfo{number}{7} (\bibinfo{year}{2022}), \bibinfo{pages}{51--55}.
\newblock


\bibitem[Banjade et~al\mbox{.}(2024)]%
        {Banjade2024EmpoweringEB}
\bibfield{author}{\bibinfo{person}{Shivraj Banjade}, \bibinfo{person}{Hiran Patel}, {and} \bibinfo{person}{Sangita Pokhrel}.} \bibinfo{year}{2024}\natexlab{}.
\newblock \showarticletitle{Empowering Education by Developing and Evaluating Generative AI-Powered Tutoring System for Enhanced Student Learning}.
\newblock \bibinfo{journal}{\emph{Journal of Artificial Intelligence and Capsule Networks}} (\bibinfo{year}{2024}).
\newblock
\urldef\tempurl%
\url{https://api.semanticscholar.org/CorpusID:272025324}
\showURL{%
\tempurl}


\bibitem[Baron(2024)]%
        {Baron2024AreAD}
\bibfield{author}{\bibinfo{person}{Philip Baron}.} \bibinfo{year}{2024}\natexlab{}.
\newblock \showarticletitle{Are AI detection and plagiarism similarity scores worthwhile in the age of ChatGPT and other Generative AI?}
\newblock \bibinfo{journal}{\emph{Scholarship of Teaching and Learning in the South}} (\bibinfo{year}{2024}).
\newblock


\bibitem[Baturay(2015)]%
        {baturay2015overview}
\bibfield{author}{\bibinfo{person}{Meltem~Huri Baturay}.} \bibinfo{year}{2015}\natexlab{}.
\newblock \showarticletitle{An overview of the world of MOOCs}.
\newblock \bibinfo{journal}{\emph{Procedia-Social and Behavioral Sciences}}  \bibinfo{volume}{174} (\bibinfo{year}{2015}), \bibinfo{pages}{427--433}.
\newblock


\bibitem[Bewersdorff et~al\mbox{.}(2025)]%
        {bewersdorff2025taking}
\bibfield{author}{\bibinfo{person}{Arne Bewersdorff}, \bibinfo{person}{Christian Hartmann}, \bibinfo{person}{Marie Hornberger}, \bibinfo{person}{Kathrin Se{\ss}ler}, \bibinfo{person}{Maria Bannert}, \bibinfo{person}{Enkelejda Kasneci}, \bibinfo{person}{Gjergji Kasneci}, \bibinfo{person}{Xiaoming Zhai}, {and} \bibinfo{person}{Claudia Nerdel}.} \bibinfo{year}{2025}\natexlab{}.
\newblock \showarticletitle{Taking the next step with generative artificial intelligence: The transformative role of multimodal large language models in science education}.
\newblock \bibinfo{journal}{\emph{Learning and Individual Differences}}  \bibinfo{volume}{118} (\bibinfo{year}{2025}), \bibinfo{pages}{102601}.
\newblock


\bibitem[Biton and Segal(2025)]%
        {Biton2025LearningTC}
\bibfield{author}{\bibinfo{person}{Yaniv Biton} {and} \bibinfo{person}{Ruti Segal}.} \bibinfo{year}{2025}\natexlab{}.
\newblock \showarticletitle{Learning to Craft and Critically Evaluate Prompts: The Role of Generative AI (ChatGPT) in Enhancing Pre-service Mathematics Teachers' TPACK and Problem-Posing Skills}.
\newblock \bibinfo{journal}{\emph{International Journal of Education in Mathematics, Science and Technology}} (\bibinfo{year}{2025}).
\newblock


\bibitem[Blok et~al\mbox{.}(2002)]%
        {blok2002computer}
\bibfield{author}{\bibinfo{person}{Henk Blok}, \bibinfo{person}{Ron Oostdam}, \bibinfo{person}{Martha~E Otter}, {and} \bibinfo{person}{Marianne Overmaat}.} \bibinfo{year}{2002}\natexlab{}.
\newblock \showarticletitle{Computer-assisted instruction in support of beginning reading instruction: A review}.
\newblock \bibinfo{journal}{\emph{Review of educational research}} \bibinfo{volume}{72}, \bibinfo{number}{1} (\bibinfo{year}{2002}), \bibinfo{pages}{101--130}.
\newblock


\bibitem[Bonner et~al\mbox{.}(2023)]%
        {Bonner2023LARGELM}
\bibfield{author}{\bibinfo{person}{Euan Bonner}, \bibinfo{person}{Ryan Lege}, {and} \bibinfo{person}{Erin Frazier}.} \bibinfo{year}{2023}\natexlab{}.
\newblock \showarticletitle{LARGE LANGUAGE MODEL-BASED ARTIFICIAL INTELLIGENCE IN THE LANGUAGE CLASSROOM: PRACTICAL IDEAS FOR TEACHING}.
\newblock \bibinfo{journal}{\emph{Teaching English With Technology}} (\bibinfo{year}{2023}).
\newblock
\urldef\tempurl%
\url{https://api.semanticscholar.org/CorpusID:256753798}
\showURL{%
\tempurl}


\bibitem[Brand{\~a}o et~al\mbox{.}(2024)]%
        {Brando2024TeacherPD}
\bibfield{author}{\bibinfo{person}{Anabela Brand{\~a}o}, \bibinfo{person}{Lu{\'i}s Pedro}, {and} \bibinfo{person}{Nelson Zagalo}.} \bibinfo{year}{2024}\natexlab{}.
\newblock \showarticletitle{Teacher professional development for a future with generative artificial intelligence – an integrative literature review}.
\newblock \bibinfo{journal}{\emph{Digital Education Review}} (\bibinfo{year}{2024}).
\newblock


\bibitem[Braun and Clarke(2019)]%
        {Braun2019ReflectingOR}
\bibfield{author}{\bibinfo{person}{Virginia Braun} {and} \bibinfo{person}{Victoria Clarke}.} \bibinfo{year}{2019}\natexlab{}.
\newblock \showarticletitle{Reflecting on reflexive thematic analysis}.
\newblock \bibinfo{journal}{\emph{Qualitative Research in Sport, Exercise and Health}}  \bibinfo{volume}{11} (\bibinfo{year}{2019}), \bibinfo{pages}{589 -- 597}.
\newblock
\urldef\tempurl%
\url{https://api.semanticscholar.org/CorpusID:197748828}
\showURL{%
\tempurl}


\bibitem[Braun and Clarke(2020)]%
        {Braun2020OneSF}
\bibfield{author}{\bibinfo{person}{Virginia Braun} {and} \bibinfo{person}{Victoria Clarke}.} \bibinfo{year}{2020}\natexlab{}.
\newblock \showarticletitle{One size fits all? What counts as quality practice in (reflexive) thematic analysis?}
\newblock \bibinfo{journal}{\emph{Qualitative Research in Psychology}}  \bibinfo{volume}{18} (\bibinfo{year}{2020}), \bibinfo{pages}{328 -- 352}.
\newblock
\urldef\tempurl%
\url{https://api.semanticscholar.org/CorpusID:225423421}
\showURL{%
\tempurl}


\bibitem[Brummelen and Lin(2020)]%
        {VanBrummelen2020EngagingTT}
\bibfield{author}{\bibinfo{person}{Jessica~Van Brummelen} {and} \bibinfo{person}{Phoebe Lin}.} \bibinfo{year}{2020}\natexlab{}.
\newblock \showarticletitle{Engaging Teachers to Co-Design Integrated AI Curriculum for K-12 Classrooms}.
\newblock \bibinfo{journal}{\emph{Proceedings of the 2021 CHI Conference on Human Factors in Computing Systems}} (\bibinfo{year}{2020}).
\newblock


\bibitem[Calo and Maclellan(2024)]%
        {calo2024towards}
\bibfield{author}{\bibinfo{person}{Tommaso Calo} {and} \bibinfo{person}{Christopher Maclellan}.} \bibinfo{year}{2024}\natexlab{}.
\newblock \showarticletitle{Towards educator-driven tutor authoring: generative AI approaches for creating intelligent tutor interfaces}. In \bibinfo{booktitle}{\emph{Proceedings of the Eleventh ACM Conference on Learning@ Scale}}. \bibinfo{pages}{305--309}.
\newblock


\bibitem[Cipolla and Lenhart(2024)]%
        {Cipolla2024GenerativeAI}
\bibfield{author}{\bibinfo{person}{B. Cipolla} {and} \bibinfo{person}{A. Lenhart}.} \bibinfo{year}{2024}\natexlab{}.
\newblock \bibinfo{booktitle}{\emph{Generative AI in K--12 Education: Challenges and Opportunities}}.
\newblock \bibinfo{type}{{T}echnical {R}eport}. \bibinfo{institution}{Common Sense Media}.
\newblock
\urldef\tempurl%
\url{https://www.commonsensemedia.org/research/generative-ai-in-k-12-education-challenges-and-opportunities}
\showURL{%
\tempurl}


\bibitem[Cuban(2001)]%
        {cuban2001oversold}
\bibfield{author}{\bibinfo{person}{Larry Cuban}.} \bibinfo{year}{2001}\natexlab{}.
\newblock \bibinfo{booktitle}{\emph{Oversold and underused: Computers in the classroom}}.
\newblock \bibinfo{publisher}{Harvard university press}.
\newblock


\bibitem[Cuban and Jandri{\'c}(2015)]%
        {cuban2015dubious}
\bibfield{author}{\bibinfo{person}{Larry Cuban} {and} \bibinfo{person}{Petar Jandri{\'c}}.} \bibinfo{year}{2015}\natexlab{}.
\newblock \showarticletitle{The dubious promise of educational technologies: Historical patterns and future challenges}.
\newblock \bibinfo{journal}{\emph{E-learning and Digital Media}} \bibinfo{volume}{12}, \bibinfo{number}{3-4} (\bibinfo{year}{2015}), \bibinfo{pages}{425--439}.
\newblock


\bibitem[Cukurova et~al\mbox{.}(2024)]%
        {cukurova2024ai}
\bibfield{author}{\bibinfo{person}{Mutlu Cukurova}, \bibinfo{person}{Fengchun Miao}, {et~al\mbox{.}}} \bibinfo{year}{2024}\natexlab{}.
\newblock \bibinfo{booktitle}{\emph{AI competency framework for teachers}}.
\newblock \bibinfo{publisher}{UNESCO Publishing}.
\newblock


\bibitem[Curwood(2014)]%
        {Curwood2014BetweenCA}
\bibfield{author}{\bibinfo{person}{Jen~Scott Curwood}.} \bibinfo{year}{2014}\natexlab{}.
\newblock \showarticletitle{Between continuity and change: identities and narratives within teacher professional development}.
\newblock \bibinfo{journal}{\emph{Teaching Education}}  \bibinfo{volume}{25} (\bibinfo{year}{2014}), \bibinfo{pages}{156 -- 183}.
\newblock


\bibitem[Dangol et~al\mbox{.}(2025)]%
        {Dangol2025AIJK}
\bibfield{author}{\bibinfo{person}{Aayushi Dangol}, \bibinfo{person}{Runhua Zhao}, \bibinfo{person}{Robert Wolfe}, \bibinfo{person}{Trushaa Ramanan}, \bibinfo{person}{Julie~A. Kientz}, {and} \bibinfo{person}{Jason~C. Yip}.} \bibinfo{year}{2025}\natexlab{}.
\newblock \showarticletitle{“AI just keeps guessing”: Using ARC Puzzles to Help Children Identify Reasoning Errors in Generative AI}.
\newblock \bibinfo{journal}{\emph{Proceedings of the 24th Interaction Design and Children}} (\bibinfo{year}{2025}).
\newblock


\bibitem[Delgado et~al\mbox{.}(2015)]%
        {delgado2015educational}
\bibfield{author}{\bibinfo{person}{Adolph~J Delgado}, \bibinfo{person}{Liane Wardlow}, \bibinfo{person}{Katherine McKnight}, {and} \bibinfo{person}{Kimberly O’Malley}.} \bibinfo{year}{2015}\natexlab{}.
\newblock \showarticletitle{Educational technology: A review of the integration, resources, and effectiveness of technology in K-12 classrooms.}
\newblock \bibinfo{journal}{\emph{Journal of Information Technology Education: Research}}  \bibinfo{volume}{14} (\bibinfo{year}{2015}).
\newblock


\bibitem[Delgado et~al\mbox{.}(2021)]%
        {Delgado2021StakeholderPI}
\bibfield{author}{\bibinfo{person}{Fernando~Pedro Delgado}, \bibinfo{person}{Stephen Yang}, \bibinfo{person}{Michael~A. Madaio}, {and} \bibinfo{person}{Qian Yang}.} \bibinfo{year}{2021}\natexlab{}.
\newblock \showarticletitle{Stakeholder Participation in AI: Beyond "Add Diverse Stakeholders and Stir"}.
\newblock \bibinfo{journal}{\emph{ArXiv}}  \bibinfo{volume}{abs/2111.01122} (\bibinfo{year}{2021}).
\newblock


\bibitem[Dhiman et~al\mbox{.}(2025)]%
        {Dhiman2025EthicalCN}
\bibfield{author}{\bibinfo{person}{Tushar Dhiman}, \bibinfo{person}{Vishakha Chauhan}, \bibinfo{person}{Asheesh Kumar}, \bibinfo{person}{M. Vasantha}, {and} \bibinfo{person}{Abhijeet Kumar}.} \bibinfo{year}{2025}\natexlab{}.
\newblock \showarticletitle{Ethical Crossroads: Navigating Data Privacy, Bias, Accountability and Sustainability in AI-Driven Education}.
\newblock \bibinfo{journal}{\emph{Open Access Journal of Multidisciplinary Research}} (\bibinfo{year}{2025}).
\newblock


\bibitem[Divanji et~al\mbox{.}(2023)]%
        {divanji2023one}
\bibfield{author}{\bibinfo{person}{Riddhi~A Divanji}, \bibinfo{person}{Samantha Bindman}, \bibinfo{person}{Allie Tung}, \bibinfo{person}{Katharine Chen}, \bibinfo{person}{Lisa Castaneda}, {and} \bibinfo{person}{Mike Scanlon}.} \bibinfo{year}{2023}\natexlab{}.
\newblock \showarticletitle{A one stop shop? Perspectives on the value of adaptive learning technologies in K-12 education}.
\newblock \bibinfo{journal}{\emph{Computers and Education Open}}  \bibinfo{volume}{5} (\bibinfo{year}{2023}), \bibinfo{pages}{100157}.
\newblock


\bibitem[{Donald J. Trump}(2025)]%
        {Trump2025AdvancingAIEducation}
\bibfield{author}{\bibinfo{person}{{Donald J. Trump}}.} \bibinfo{year}{2025}\natexlab{}.
\newblock \bibinfo{title}{Advancing Artificial Intelligence Education for American Youth}.
\newblock \bibinfo{howpublished}{Executive Order, The White House}.
\newblock
\urldef\tempurl%
\url{https://www.whitehouse.gov/presidential-actions/2025/04/advancing-artificial-intelligence-education-for-american-youth/}
\showURL{%
\tempurl}
\newblock
\shownote{Accessed: 2025-09-02}.


\bibitem[Druga et~al\mbox{.}(2021)]%
        {druga20214as}
\bibfield{author}{\bibinfo{person}{Stefania Druga}, \bibinfo{person}{Jason Yip}, \bibinfo{person}{Michael Preston}, {and} \bibinfo{person}{Devin Dillon}.} \bibinfo{year}{2021}\natexlab{}.
\newblock \showarticletitle{The 4As: Ask, adapt, author, analyze-AI literacy framework for families}.
\newblock \bibinfo{journal}{\emph{Algorithmic Rights and Protections for Children}} (\bibinfo{year}{2021}).
\newblock


\bibitem[Ellis and Slade(2023)]%
        {ellis2023new}
\bibfield{author}{\bibinfo{person}{Amanda~R Ellis} {and} \bibinfo{person}{Emily Slade}.} \bibinfo{year}{2023}\natexlab{}.
\newblock \showarticletitle{A new era of learning: Considerations for ChatGPT as a tool to enhance statistics and data science education}.
\newblock \bibinfo{journal}{\emph{Journal of Statistics and Data Science Education}} \bibinfo{volume}{31}, \bibinfo{number}{2} (\bibinfo{year}{2023}), \bibinfo{pages}{128--133}.
\newblock


\bibitem[Elsayary et~al\mbox{.}(2025)]%
        {Elsayary2025ExaminingTR}
\bibfield{author}{\bibinfo{person}{Areej Elsayary}, \bibinfo{person}{Mohammad~Amin Kuhail}, {and} \bibinfo{person}{Zeina Hojeij}.} \bibinfo{year}{2025}\natexlab{}.
\newblock \showarticletitle{Examining the Role of Prompt Engineering in Utilizing Generative AI Tools for Lesson Planning: Insights From Teachers’ Experiences and Perceptions}.
\newblock \bibinfo{journal}{\emph{Human Behavior and Emerging Technologies}} (\bibinfo{year}{2025}).
\newblock


\bibitem[Ferman et~al\mbox{.}(2021)]%
        {Ferman2021ArtificialIT}
\bibfield{author}{\bibinfo{person}{Bruno Ferman}, \bibinfo{person}{Lycia Lima}, {and} \bibinfo{person}{Flavio Luiz~Russo Riva}.} \bibinfo{year}{2021}\natexlab{}.
\newblock \showarticletitle{Artificial Intelligence, Teacher Tasks and Individualized Pedagogy}.
\newblock


\bibitem[Floris et~al\mbox{.}(2024)]%
        {Floris2024ArtificialII}
\bibfield{author}{\bibinfo{person}{Flora~Debora Floris}, \bibinfo{person}{Utami Widiati}, \bibinfo{person}{Willy~Ardian Renandya}, {and} \bibinfo{person}{Yazid Basthomi}.} \bibinfo{year}{2024}\natexlab{}.
\newblock \showarticletitle{Artificial Intelligence in English Language Teaching: Fostering Joint Enterprise in Online Communities}.
\newblock \bibinfo{journal}{\emph{JEES (Journal of English Educators Society)}} (\bibinfo{year}{2024}).
\newblock


\bibitem[Forte et~al\mbox{.}(2012)]%
        {Forte2012GrassrootsPD}
\bibfield{author}{\bibinfo{person}{Andrea Forte}, \bibinfo{person}{Melissa Humphreys}, {and} \bibinfo{person}{Thomas~H. Park}.} \bibinfo{year}{2012}\natexlab{}.
\newblock \showarticletitle{Grassroots Professional Development: How Teachers Use Twitter}.
\newblock \bibinfo{journal}{\emph{Proceedings of the International AAAI Conference on Web and Social Media}} (\bibinfo{year}{2012}).
\newblock
\urldef\tempurl%
\url{https://api.semanticscholar.org/CorpusID:15426349}
\showURL{%
\tempurl}


\bibitem[Gabriel(2024)]%
        {Gabriel2024GenerativeAA}
\bibfield{author}{\bibinfo{person}{Sonja Gabriel}.} \bibinfo{year}{2024}\natexlab{}.
\newblock \showarticletitle{Generative AI and Educational (In)Equity}.
\newblock \bibinfo{journal}{\emph{International Conference on AI Research}} (\bibinfo{year}{2024}).
\newblock


\bibitem[Ghimire and Edwards(2024a)]%
        {ghimire2024coding}
\bibfield{author}{\bibinfo{person}{Aashish Ghimire} {and} \bibinfo{person}{John Edwards}.} \bibinfo{year}{2024}\natexlab{a}.
\newblock \showarticletitle{Coding with ai: How are tools like chatgpt being used by students in foundational programming courses}. In \bibinfo{booktitle}{\emph{International Conference on Artificial Intelligence in Education}}. Springer, \bibinfo{pages}{259--267}.
\newblock


\bibitem[Ghimire and Edwards(2024b)]%
        {Ghimire2024FromGT}
\bibfield{author}{\bibinfo{person}{Aashish Ghimire} {and} \bibinfo{person}{John Edwards}.} \bibinfo{year}{2024}\natexlab{b}.
\newblock \showarticletitle{From Guidelines to Governance: A Study of AI Policies in Education}. In \bibinfo{booktitle}{\emph{AIED Companion}}.
\newblock
\urldef\tempurl%
\url{https://api.semanticscholar.org/CorpusID:268680559}
\showURL{%
\tempurl}


\bibitem[Goldstein(2025)]%
        {Goldstein2025TeachersAI}
\bibfield{author}{\bibinfo{person}{Dana Goldstein}.} \bibinfo{year}{2025}\natexlab{}.
\newblock \bibinfo{booktitle}{\emph{Teachers Worry About Students Using A.I. But They Love It for Themselves}}.
\newblock
\urldef\tempurl%
\url{https://www.nytimes.com/2025/04/14/us/schools-ai-teachers-writing.html}
\showURL{%
\tempurl}
\newblock
\shownote{Accessed: 2025-09-02}.


\bibitem[Graesser et~al\mbox{.}(2012)]%
        {graesser2012intelligent}
\bibfield{author}{\bibinfo{person}{Arthur~C Graesser}, \bibinfo{person}{Mark~W Conley}, {and} \bibinfo{person}{Andrew Olney}.} \bibinfo{year}{2012}\natexlab{}.
\newblock \showarticletitle{Intelligent tutoring systems.}
\newblock  (\bibinfo{year}{2012}).
\newblock


\bibitem[Gu(2014)]%
        {Gu2014TheRO}
\bibfield{author}{\bibinfo{person}{Qing Gu}.} \bibinfo{year}{2014}\natexlab{}.
\newblock \showarticletitle{The role of relational resilience in teachers’ career-long commitment and effectiveness}.
\newblock \bibinfo{journal}{\emph{Teachers and Teaching}}  \bibinfo{volume}{20} (\bibinfo{year}{2014}), \bibinfo{pages}{502 -- 529}.
\newblock


\bibitem[Han et~al\mbox{.}(2024)]%
        {Han2024TeachersPA}
\bibfield{author}{\bibinfo{person}{Ariel Han}, \bibinfo{person}{Xiaofei Zhou}, \bibinfo{person}{Zhenyao Cai}, \bibinfo{person}{Shenshen Han}, \bibinfo{person}{Richard Ko}, \bibinfo{person}{Seth Corrigan}, {and} \bibinfo{person}{Kylie~A Peppler}.} \bibinfo{year}{2024}\natexlab{}.
\newblock \showarticletitle{Teachers, Parents, and Students' perspectives on Integrating Generative AI into Elementary Literacy Education}.
\newblock \bibinfo{journal}{\emph{Proceedings of the 2024 CHI Conference on Human Factors in Computing Systems}} (\bibinfo{year}{2024}).
\newblock


\bibitem[He et~al\mbox{.}(2025)]%
        {he2025carlitos}
\bibfield{author}{\bibinfo{person}{Kunlei He}, \bibinfo{person}{Xuechen Liu}, \bibinfo{person}{Ying Xu}, \bibinfo{person}{Andres~S Bustamante}, {and} \bibinfo{person}{Mark Warschauer}.} \bibinfo{year}{2025}\natexlab{}.
\newblock \showarticletitle{“Carlitos the Curious Caterpillar”: Exploring Teacher-AI Co-Creation of Culturally Responsive Educational Materials for Young Learners}.
\newblock In \bibinfo{booktitle}{\emph{Proceedings of the 24th Interaction Design and Children}}. \bibinfo{pages}{236--254}.
\newblock


\bibitem[Henriksen et~al\mbox{.}(2024)]%
        {Henriksen2024CreativeLF}
\bibfield{author}{\bibinfo{person}{Danah Henriksen}, \bibinfo{person}{Punya Mishra}, {and} \bibinfo{person}{Rachel Stern}.} \bibinfo{year}{2024}\natexlab{}.
\newblock \showarticletitle{Creative Learning for Sustainability in a World of AI: Action, Mindset, Values}.
\newblock \bibinfo{journal}{\emph{Sustainability}} (\bibinfo{year}{2024}).
\newblock
\urldef\tempurl%
\url{https://api.semanticscholar.org/CorpusID:270032473}
\showURL{%
\tempurl}


\bibitem[Hillary Greene~Nolan et~al\mbox{.}(2024)]%
        {GreeneNolanPhD2024TeachingPG}
\bibfield{author}{\bibinfo{person}{Ph.D. Hillary Greene~Nolan}, \bibinfo{person}{Ph.D. Merijke~Coenraad}, {and} \bibinfo{person}{Ph.D. Viki M.~Young}.} \bibinfo{year}{2024}\natexlab{}.
\newblock \showarticletitle{Teaching Partner, Grading Assistant, Substitute Teacher: Three Ways Teachers Positioned an Artificial Intelligence Tool in Writing Instruction}.
\newblock


\bibitem[Holstein et~al\mbox{.}(2019)]%
        {Holstein2019DesigningFC}
\bibfield{author}{\bibinfo{person}{Kenneth Holstein}, \bibinfo{person}{Bruce~M. McLaren}, {and} \bibinfo{person}{Vincent Aleven}.} \bibinfo{year}{2019}\natexlab{}.
\newblock \showarticletitle{Designing for Complementarity: Teacher and Student Needs for Orchestration Support in AI-Enhanced Classrooms}. In \bibinfo{booktitle}{\emph{International Conference on Artificial Intelligence in Education}}.
\newblock


\bibitem[Hu et~al\mbox{.}(2025)]%
        {Hu2025GenerativeAI}
\bibfield{author}{\bibinfo{person}{Xiangen Hu}, \bibinfo{person}{Sheng Xu}, \bibinfo{person}{Richard~Jiarui Tong}, {and} \bibinfo{person}{Art Graesser}.} \bibinfo{year}{2025}\natexlab{}.
\newblock \showarticletitle{Generative AI in Education: From Foundational Insights to the Socratic Playground for Learning}.
\newblock \bibinfo{journal}{\emph{ArXiv}}  \bibinfo{volume}{abs/2501.06682} (\bibinfo{year}{2025}).
\newblock
\urldef\tempurl%
\url{https://api.semanticscholar.org/CorpusID:275470916}
\showURL{%
\tempurl}


\bibitem[Impey et~al\mbox{.}(2024)]%
        {Impey2024UsingLL}
\bibfield{author}{\bibinfo{person}{Christopher Impey}, \bibinfo{person}{Matthew~C. Wenger}, \bibinfo{person}{Nikhil Garuda}, \bibinfo{person}{Shahriar Golchin}, {and} \bibinfo{person}{Sarah Stamer}.} \bibinfo{year}{2024}\natexlab{}.
\newblock \showarticletitle{Using Large Language Models for Automated Grading of Student Writing about Science}.
\newblock \bibinfo{journal}{\emph{ArXiv}}  \bibinfo{volume}{abs/2412.18719} (\bibinfo{year}{2024}).
\newblock
\urldef\tempurl%
\url{https://api.semanticscholar.org/CorpusID:275119013}
\showURL{%
\tempurl}


\bibitem[{Indiana Department of Education}(2024)]%
        {IndianaDOE2024AIReport}
\bibfield{author}{\bibinfo{person}{{Indiana Department of Education}}.} \bibinfo{year}{2024}\natexlab{}.
\newblock \bibinfo{booktitle}{\emph{AI-Powered Platform Pilot Grant Final Report}}.
\newblock \bibinfo{type}{{T}echnical {R}eport}. \bibinfo{institution}{Indiana Department of Education}, \bibinfo{address}{Indianapolis, IN}.
\newblock
\newblock
\shownote{Pilot grant final report for 2023--2024}.


\bibitem[Jin et~al\mbox{.}(2025)]%
        {Jin2025KnowledgeA}
\bibfield{author}{\bibinfo{person}{Fangzhou Jin}, \bibinfo{person}{Xiangmei Peng}, \bibinfo{person}{Lanfang Sun}, \bibinfo{person}{Zicong Song}, \bibinfo{person}{Keyi Zhou}, {and} \bibinfo{person}{Chin-Hsi Lin}.} \bibinfo{year}{2025}\natexlab{}.
\newblock \showarticletitle{Knowledge (Co-)Construction Among Artificial Intelligence, Novice Teachers, and Experienced Teachers in an Online Professional Learning Community}.
\newblock \bibinfo{journal}{\emph{J. Comput. Assist. Learn.}}  \bibinfo{volume}{41} (\bibinfo{year}{2025}).
\newblock
\urldef\tempurl%
\url{https://api.semanticscholar.org/CorpusID:276303471}
\showURL{%
\tempurl}


\bibitem[Johnson et~al\mbox{.}(2016)]%
        {Johnson2016ChallengesAS}
\bibfield{author}{\bibinfo{person}{Amy~M. Johnson}, \bibinfo{person}{Matthew~E. Jacovina}, \bibinfo{person}{Devin~G. Russell}, {and} \bibinfo{person}{Christian~M. Soto}.} \bibinfo{year}{2016}\natexlab{}.
\newblock \showarticletitle{Challenges and Solutions When Using Technologies in the Classroom.}
\newblock
\urldef\tempurl%
\url{https://api.semanticscholar.org/CorpusID:217791189}
\showURL{%
\tempurl}


\bibitem[Joshi(2025)]%
        {Joshi2025StrategicIO}
\bibfield{author}{\bibinfo{person}{Satyadhar Joshi}.} \bibinfo{year}{2025}\natexlab{}.
\newblock \showarticletitle{Strategic Integration of Artificial Intelligence in U.S. K–12 Education: A Comprehensive Review and Policy Roadmap}.
\newblock \bibinfo{journal}{\emph{International Journal of Computer Applications}} (\bibinfo{year}{2025}).
\newblock
\urldef\tempurl%
\url{https://api.semanticscholar.org/CorpusID:280043529}
\showURL{%
\tempurl}


\bibitem[Kaghan and Bowker(2001)]%
        {kaghan2001out}
\bibfield{author}{\bibinfo{person}{William~N Kaghan} {and} \bibinfo{person}{Geoffrey~C Bowker}.} \bibinfo{year}{2001}\natexlab{}.
\newblock \showarticletitle{Out of machine age?: complexity, sociotechnical systems and actor network theory}.
\newblock \bibinfo{journal}{\emph{Journal of Engineering and Technology Management}} \bibinfo{volume}{18}, \bibinfo{number}{3-4} (\bibinfo{year}{2001}), \bibinfo{pages}{253--269}.
\newblock


\bibitem[Kang et~al\mbox{.}(2025)]%
        {kang2025tutorcraftease}
\bibfield{author}{\bibinfo{person}{Wenhui Kang}, \bibinfo{person}{Lin Zhang}, \bibinfo{person}{Xiaolan Peng}, \bibinfo{person}{Hao Zhang}, \bibinfo{person}{Anchi Li}, \bibinfo{person}{Mengyao Wang}, \bibinfo{person}{Jin Huang}, \bibinfo{person}{Feng Tian}, {and} \bibinfo{person}{Guozhong Dai}.} \bibinfo{year}{2025}\natexlab{}.
\newblock \showarticletitle{TutorCraftEase: Enhancing Pedagogical Question Creation with Large Language Models}. In \bibinfo{booktitle}{\emph{Proceedings of the 2025 CHI Conference on Human Factors in Computing Systems}}. \bibinfo{pages}{1--22}.
\newblock


\bibitem[Kem(2022)]%
        {kem2022personalised}
\bibfield{author}{\bibinfo{person}{Deepak Kem}.} \bibinfo{year}{2022}\natexlab{}.
\newblock \showarticletitle{Personalised and adaptive learning: Emerging learning platforms in the era of digital and smart learning}.
\newblock \bibinfo{journal}{\emph{International Journal of Social Science and Human Research}} \bibinfo{volume}{5}, \bibinfo{number}{2} (\bibinfo{year}{2022}), \bibinfo{pages}{385--391}.
\newblock


\bibitem[Kim(2025)]%
        {Kim2025PerceptionsAP}
\bibfield{author}{\bibinfo{person}{Juhee Kim}.} \bibinfo{year}{2025}\natexlab{}.
\newblock \showarticletitle{Perceptions and preparedness of K-12 educators in adopting generative AI}.
\newblock \bibinfo{journal}{\emph{Research in Learning Technology}} (\bibinfo{year}{2025}).
\newblock
\urldef\tempurl%
\url{https://api.semanticscholar.org/CorpusID:279681892}
\showURL{%
\tempurl}


\bibitem[Koh and Doroudi(2023)]%
        {koh2023learning}
\bibfield{author}{\bibinfo{person}{Elizabeth Koh} {and} \bibinfo{person}{Shayan Doroudi}.} \bibinfo{year}{2023}\natexlab{}.
\newblock \bibinfo{title}{Learning, teaching, and assessment with generative artificial intelligence: towards a plateau of productivity}.
\newblock , \bibinfo{numpages}{109--116}~pages.
\newblock


\bibitem[Kohnke et~al\mbox{.}(2023)]%
        {Kohnke2023ExploringGA}
\bibfield{author}{\bibinfo{person}{Lucas Kohnke}, \bibinfo{person}{Benjamin~Luke Moorhouse}, {and} \bibinfo{person}{Di Zou}.} \bibinfo{year}{2023}\natexlab{}.
\newblock \showarticletitle{Exploring generative artificial intelligence preparedness among university language instructors: A case study}.
\newblock \bibinfo{journal}{\emph{Comput. Educ. Artif. Intell.}}  \bibinfo{volume}{5} (\bibinfo{year}{2023}), \bibinfo{pages}{100156}.
\newblock


\bibitem[Kohnke et~al\mbox{.}(2024)]%
        {kohnke2024technostress}
\bibfield{author}{\bibinfo{person}{Lucas Kohnke}, \bibinfo{person}{Di Zou}, {and} \bibinfo{person}{Benjamin~L Moorhouse}.} \bibinfo{year}{2024}\natexlab{}.
\newblock \showarticletitle{Technostress and English language teaching in the age of generative AI}.
\newblock \bibinfo{journal}{\emph{Educational technology \& society}} \bibinfo{volume}{27}, \bibinfo{number}{2} (\bibinfo{year}{2024}), \bibinfo{pages}{306--320}.
\newblock


\bibitem[Kong and Yang(2024)]%
        {Kong2024AHL}
\bibfield{author}{\bibinfo{person}{Siu-Cheung Kong} {and} \bibinfo{person}{Yin Yang}.} \bibinfo{year}{2024}\natexlab{}.
\newblock \showarticletitle{A Human-Centered Learning and Teaching Framework Using Generative Artificial Intelligence for Self-Regulated Learning Development Through Domain Knowledge Learning in K–12 Settings}.
\newblock \bibinfo{journal}{\emph{IEEE Transactions on Learning Technologies}}  \bibinfo{volume}{17} (\bibinfo{year}{2024}), \bibinfo{pages}{1588--1599}.
\newblock
\urldef\tempurl%
\url{https://api.semanticscholar.org/CorpusID:269343653}
\showURL{%
\tempurl}


\bibitem[Kopcha et~al\mbox{.}(2020)]%
        {kopcha2020process}
\bibfield{author}{\bibinfo{person}{Theodore~J Kopcha}, \bibinfo{person}{Kalianne~L Neumann}, \bibinfo{person}{Anne Ottenbreit-Leftwich}, {and} \bibinfo{person}{Elizabeth Pitman}.} \bibinfo{year}{2020}\natexlab{}.
\newblock \showarticletitle{Process over product: The next evolution of our quest for technology integration}.
\newblock \bibinfo{journal}{\emph{Educational Technology Research and Development}} \bibinfo{volume}{68}, \bibinfo{number}{2} (\bibinfo{year}{2020}), \bibinfo{pages}{729--749}.
\newblock


\bibitem[Kozma(2003)]%
        {Kozma2003TechnologyAC}
\bibfield{author}{\bibinfo{person}{Robert Kozma}.} \bibinfo{year}{2003}\natexlab{}.
\newblock \showarticletitle{Technology and Classroom Practices}.
\newblock \bibinfo{journal}{\emph{Journal of Research on Technology in Education}}  \bibinfo{volume}{36} (\bibinfo{year}{2003}), \bibinfo{pages}{1 -- 14}.
\newblock


\bibitem[Laak and Aru(2024)]%
        {laak2024generative}
\bibfield{author}{\bibinfo{person}{Kristjan-Julius Laak} {and} \bibinfo{person}{Jaan Aru}.} \bibinfo{year}{2024}\natexlab{}.
\newblock \showarticletitle{Generative AI in K-12: Opportunities for learning and utility for teachers}. In \bibinfo{booktitle}{\emph{International conference on artificial intelligence in education}}. Springer, \bibinfo{pages}{502--509}.
\newblock


\bibitem[Laato et~al\mbox{.}(2023)]%
        {Laato2023AIAssistedLW}
\bibfield{author}{\bibinfo{person}{Samuli Laato}, \bibinfo{person}{Benedikt Morschheuser}, \bibinfo{person}{Juho Hamari}, {and} \bibinfo{person}{Jari Bj{\"o}rne}.} \bibinfo{year}{2023}\natexlab{}.
\newblock \showarticletitle{AI-Assisted Learning with ChatGPT and Large Language Models: Implications for Higher Education}.
\newblock \bibinfo{journal}{\emph{2023 IEEE International Conference on Advanced Learning Technologies (ICALT)}} (\bibinfo{year}{2023}), \bibinfo{pages}{226--230}.
\newblock
\urldef\tempurl%
\url{https://api.semanticscholar.org/CorpusID:263228794}
\showURL{%
\tempurl}


\bibitem[Latour(1987)]%
        {latour1987science}
\bibfield{author}{\bibinfo{person}{Bruno Latour}.} \bibinfo{year}{1987}\natexlab{}.
\newblock \bibinfo{booktitle}{\emph{Science in action: How to follow scientists and engineers through society}}.
\newblock \bibinfo{publisher}{Harvard university press}.
\newblock


\bibitem[Lawrie(2023)]%
        {Lawrie2023EstablishingAD}
\bibfield{author}{\bibinfo{person}{Gwendolyn~Angela Lawrie}.} \bibinfo{year}{2023}\natexlab{}.
\newblock \showarticletitle{Establishing a delicate balance in the relationship between artificial intelligence and authentic assessment in student learning}.
\newblock \bibinfo{journal}{\emph{Chemistry Education Research and Practice}} (\bibinfo{year}{2023}).
\newblock


\bibitem[Lee et~al\mbox{.}(2024)]%
        {Lee2024CheatingIT}
\bibfield{author}{\bibinfo{person}{Victor~R. Lee}, \bibinfo{person}{Denise Pope}, \bibinfo{person}{Sarah Miles}, {and} \bibinfo{person}{Rosalia Zarate}.} \bibinfo{year}{2024}\natexlab{}.
\newblock \showarticletitle{Cheating in the age of generative AI: A high school survey study of cheating behaviors before and after the release of ChatGPT}.
\newblock \bibinfo{journal}{\emph{Comput. Educ. Artif. Intell.}}  \bibinfo{volume}{7} (\bibinfo{year}{2024}), \bibinfo{pages}{100253}.
\newblock


\bibitem[Leonardi(2009)]%
        {leonardi2009crossing}
\bibfield{author}{\bibinfo{person}{Paul~M Leonardi}.} \bibinfo{year}{2009}\natexlab{}.
\newblock \showarticletitle{Crossing the implementation line: The mutual constitution of technology and organizing across development and use activities}.
\newblock \bibinfo{journal}{\emph{Communication Theory}} \bibinfo{volume}{19}, \bibinfo{number}{3} (\bibinfo{year}{2009}), \bibinfo{pages}{278--310}.
\newblock


\bibitem[Lewis et~al\mbox{.}(2025)]%
        {lewis2025exploring}
\bibfield{author}{\bibinfo{person}{Aaleyah Lewis}, \bibinfo{person}{Aayushi Dangol}, \bibinfo{person}{Hyewon Suh}, \bibinfo{person}{Abbie Olszewski}, \bibinfo{person}{James Fogarty}, {and} \bibinfo{person}{Julie~A Kientz}.} \bibinfo{year}{2025}\natexlab{}.
\newblock \showarticletitle{Exploring AI-Based Support in Speech-Language Pathology for Culturally and Linguistically Diverse Children}. In \bibinfo{booktitle}{\emph{Proceedings of the 2025 CHI Conference on Human Factors in Computing Systems}}. \bibinfo{pages}{1--19}.
\newblock


\bibitem[Li et~al\mbox{.}(2024)]%
        {Li2024AutomateKC}
\bibfield{author}{\bibinfo{person}{Hang Li}, \bibinfo{person}{Tianlong Xu}, \bibinfo{person}{Jiliang Tang}, {and} \bibinfo{person}{Qingsong Wen}.} \bibinfo{year}{2024}\natexlab{}.
\newblock \showarticletitle{Automate Knowledge Concept Tagging on Math Questions with LLMs}.
\newblock \bibinfo{journal}{\emph{ArXiv}}  \bibinfo{volume}{abs/2403.17281} (\bibinfo{year}{2024}).
\newblock
\urldef\tempurl%
\url{https://api.semanticscholar.org/CorpusID:268691366}
\showURL{%
\tempurl}


\bibitem[Lin et~al\mbox{.}(2024)]%
        {lin2024s}
\bibfield{author}{\bibinfo{person}{Luona Lin}, \bibinfo{person}{Kim Parker}, {and} \bibinfo{person}{Juliana Horowitz}.} \bibinfo{year}{2024}\natexlab{}.
\newblock \showarticletitle{What's It Like to Be a Teacher in America Today?.}
\newblock \bibinfo{journal}{\emph{Pew Research Center}} (\bibinfo{year}{2024}).
\newblock


\bibitem[Lin et~al\mbox{.}(2020)]%
        {lin2020zhorai}
\bibfield{author}{\bibinfo{person}{Phoebe Lin}, \bibinfo{person}{Jessica Van~Brummelen}, \bibinfo{person}{Galit Lukin}, \bibinfo{person}{Randi Williams}, {and} \bibinfo{person}{Cynthia Breazeal}.} \bibinfo{year}{2020}\natexlab{}.
\newblock \showarticletitle{Zhorai: Designing a conversational agent for children to explore machine learning concepts}. In \bibinfo{booktitle}{\emph{Proceedings of the AAAI conference on artificial intelligence}}, Vol.~\bibinfo{volume}{34}. \bibinfo{pages}{13381--13388}.
\newblock


\bibitem[Long and Magerko(2020)]%
        {Long2020WhatIA}
\bibfield{author}{\bibinfo{person}{Duri Long} {and} \bibinfo{person}{Brian Magerko}.} \bibinfo{year}{2020}\natexlab{}.
\newblock \showarticletitle{What is AI Literacy? Competencies and Design Considerations}.
\newblock \bibinfo{journal}{\emph{Proceedings of the 2020 CHI Conference on Human Factors in Computing Systems}} (\bibinfo{year}{2020}).
\newblock
\urldef\tempurl%
\url{https://api.semanticscholar.org/CorpusID:211264278}
\showURL{%
\tempurl}


\bibitem[Lu et~al\mbox{.}(2024)]%
        {lu2024supporting}
\bibfield{author}{\bibinfo{person}{Jijian Lu}, \bibinfo{person}{Ruxin Zheng}, \bibinfo{person}{Zikun Gong}, {and} \bibinfo{person}{Huifen Xu}.} \bibinfo{year}{2024}\natexlab{}.
\newblock \showarticletitle{Supporting teachers’ professional development with generative AI: The effects on higher order thinking and self-efficacy}.
\newblock \bibinfo{journal}{\emph{IEEE Transactions on Learning Technologies}}  \bibinfo{volume}{17} (\bibinfo{year}{2024}), \bibinfo{pages}{1267--1277}.
\newblock


\bibitem[Maas and Ingl{\'e}s(2024)]%
        {Maas2024BeyondPA}
\bibfield{author}{\bibinfo{person}{Jonne Maas} {and} \bibinfo{person}{Aar{\'o}n~Moreno Ingl{\'e}s}.} \bibinfo{year}{2024}\natexlab{}.
\newblock \showarticletitle{Beyond Participatory AI}. In \bibinfo{booktitle}{\emph{AAAI/ACM Conference on AI, Ethics, and Society}}.
\newblock
\urldef\tempurl%
\url{https://api.semanticscholar.org/CorpusID:274243377}
\showURL{%
\tempurl}


\bibitem[MacDowell et~al\mbox{.}(2024)]%
        {macdowell2024preparing}
\bibfield{author}{\bibinfo{person}{Paula MacDowell}, \bibinfo{person}{Kristin Moskalyk}, \bibinfo{person}{Katrina Korchinski}, {and} \bibinfo{person}{Dirk Morrison}.} \bibinfo{year}{2024}\natexlab{}.
\newblock \showarticletitle{Preparing educators to teach and create with generative artificial intelligence}.
\newblock \bibinfo{journal}{\emph{Canadian Journal of Learning and Technology}} \bibinfo{volume}{50}, \bibinfo{number}{4} (\bibinfo{year}{2024}), \bibinfo{pages}{1--23}.
\newblock


\bibitem[Maceli et~al\mbox{.}(2024)]%
        {Maceli2024IncorporatingUU}
\bibfield{author}{\bibinfo{person}{Monica Maceli}, \bibinfo{person}{Nancy Smith}, {and} \bibinfo{person}{Gatha Bhakta}.} \bibinfo{year}{2024}\natexlab{}.
\newblock \showarticletitle{Incorporating Unanticipated Uses of Generative AI into HCI Education}.
\newblock \bibinfo{journal}{\emph{Proceedings of the 6th Annual Symposium on HCI Education}} (\bibinfo{year}{2024}).
\newblock
\urldef\tempurl%
\url{https://api.semanticscholar.org/CorpusID:270129751}
\showURL{%
\tempurl}


\bibitem[Mah et~al\mbox{.}(2024)]%
        {Mah2024BeyondCE}
\bibfield{author}{\bibinfo{person}{Chris Mah}, \bibinfo{person}{Hillary Walker}, \bibinfo{person}{Lena Phalen}, \bibinfo{person}{Sarah Levine}, \bibinfo{person}{Sarah~W. Beck}, {and} \bibinfo{person}{Jaylen Pittman}.} \bibinfo{year}{2024}\natexlab{}.
\newblock \showarticletitle{Beyond CheatBots: Examining Tensions in Teachers’ and Students’ Perceptions of Cheating and Learning with ChatGPT}.
\newblock \bibinfo{journal}{\emph{Education Sciences}} (\bibinfo{year}{2024}).
\newblock


\bibitem[Maity and Deroy(2024)]%
        {Maity2024TheFO}
\bibfield{author}{\bibinfo{person}{Subhankar Maity} {and} \bibinfo{person}{Aniket Deroy}.} \bibinfo{year}{2024}\natexlab{}.
\newblock \showarticletitle{The Future of Learning in the Age of Generative AI: Automated Question Generation and Assessment with Large Language Models}.
\newblock \bibinfo{journal}{\emph{ArXiv}}  \bibinfo{volume}{abs/2410.09576} (\bibinfo{year}{2024}).
\newblock
\urldef\tempurl%
\url{https://api.semanticscholar.org/CorpusID:273346232}
\showURL{%
\tempurl}


\bibitem[Mollick et~al\mbox{.}(2024)]%
        {mollick2024ai}
\bibfield{author}{\bibinfo{person}{Ethan Mollick}, \bibinfo{person}{Lilach Mollick}, \bibinfo{person}{Natalie Bach}, \bibinfo{person}{LJ Ciccarelli}, \bibinfo{person}{Ben Przystanski}, {and} \bibinfo{person}{Daniel Ravipinto}.} \bibinfo{year}{2024}\natexlab{}.
\newblock \showarticletitle{AI agents and education: Simulated practice at scale}.
\newblock \bibinfo{journal}{\emph{arXiv preprint arXiv:2407.12796}} (\bibinfo{year}{2024}).
\newblock


\bibitem[Moorhouse et~al\mbox{.}(2023)]%
        {moorhouse2023generative}
\bibfield{author}{\bibinfo{person}{Benjamin~Luke Moorhouse}, \bibinfo{person}{Marie~Alina Yeo}, {and} \bibinfo{person}{Yuwei Wan}.} \bibinfo{year}{2023}\natexlab{}.
\newblock \showarticletitle{Generative AI tools and assessment: Guidelines of the world's top-ranking universities}.
\newblock \bibinfo{journal}{\emph{Computers and Education Open}}  \bibinfo{volume}{5} (\bibinfo{year}{2023}), \bibinfo{pages}{100151}.
\newblock


\bibitem[Morales-Navarro et~al\mbox{.}(2024)]%
        {morales2024s}
\bibfield{author}{\bibinfo{person}{Luis Morales-Navarro}, \bibinfo{person}{Phillip Gao}, \bibinfo{person}{Eric Yang}, {and} \bibinfo{person}{Yasmin~B Kafai}.} \bibinfo{year}{2024}\natexlab{}.
\newblock \showarticletitle{" It's smart and it's stupid:" Youth's conflicting perspectives on LLMs' language comprehension and ethics}. In \bibinfo{booktitle}{\emph{Proceedings of the 19th WiPSCE Conference on Primary and Secondary Computing Education Research}}. \bibinfo{pages}{1--2}.
\newblock


\bibitem[Ng et~al\mbox{.}(2025)]%
        {ng2025opportunities}
\bibfield{author}{\bibinfo{person}{Davy Tsz~Kit Ng}, \bibinfo{person}{Eagle Kai~Chi Chan}, {and} \bibinfo{person}{Chung~Kwan Lo}.} \bibinfo{year}{2025}\natexlab{}.
\newblock \showarticletitle{Opportunities, challenges and school strategies for integrating generative AI in education}.
\newblock \bibinfo{journal}{\emph{Computers and Education: Artificial Intelligence}} (\bibinfo{year}{2025}), \bibinfo{pages}{100373}.
\newblock


\bibitem[Novita(2025)]%
        {Novita2025AIIL}
\bibfield{author}{\bibinfo{person}{Rian Novita}.} \bibinfo{year}{2025}\natexlab{}.
\newblock \showarticletitle{AI in Lesson Planning: Improving Teacher Efficiency and Instructional Design}.
\newblock \bibinfo{journal}{\emph{JURNAL RISET RUMPUN ILMU PENDIDIKAN}} (\bibinfo{year}{2025}).
\newblock


\bibitem[of~Educational~Technology(2023)]%
        {office2023artificial}
\bibfield{author}{\bibinfo{person}{Office of Educational~Technology}.} \bibinfo{year}{2023}\natexlab{}.
\newblock \bibinfo{title}{Artificial intelligence and the future of teaching and learning: Insights and recommendations}.
\newblock
\newblock


\bibitem[Palinkas et~al\mbox{.}(2015)]%
        {Palinkas2015PurposefulSF}
\bibfield{author}{\bibinfo{person}{Lawrence~A. Palinkas}, \bibinfo{person}{Sarah~McCue Horwitz}, \bibinfo{person}{Carla~A Green}, \bibinfo{person}{Jennifer~P. Wisdom}, \bibinfo{person}{Naihua Duan}, {and} \bibinfo{person}{Kimberly~Eaton Hoagwood}.} \bibinfo{year}{2015}\natexlab{}.
\newblock \showarticletitle{Purposeful Sampling for Qualitative Data Collection and Analysis in Mixed Method Implementation Research}.
\newblock \bibinfo{journal}{\emph{Administration and Policy in Mental Health and Mental Health Services Research}}  \bibinfo{volume}{42} (\bibinfo{year}{2015}), \bibinfo{pages}{533--544}.
\newblock
\urldef\tempurl%
\url{https://api.semanticscholar.org/CorpusID:16064662}
\showURL{%
\tempurl}


\bibitem[Payne(2019)]%
        {payne2019ethics}
\bibfield{author}{\bibinfo{person}{Blakeley~H Payne}.} \bibinfo{year}{2019}\natexlab{}.
\newblock \showarticletitle{An ethics of artificial intelligence curriculum for middle school students}.
\newblock \bibinfo{journal}{\emph{MIT Media Lab Personal Robots Group. Retrieved Oct}}  \bibinfo{volume}{10} (\bibinfo{year}{2019}), \bibinfo{pages}{2019}.
\newblock


\bibitem[Prather et~al\mbox{.}(2024)]%
        {Prather2024BeyondTH}
\bibfield{author}{\bibinfo{person}{James Prather}, \bibinfo{person}{Juho Leinonen}, \bibinfo{person}{Natalie Kiesler}, \bibinfo{person}{Jamie~Gorson Benario}, \bibinfo{person}{Sam Lau}, \bibinfo{person}{Stephen Macneil}, \bibinfo{person}{Narges Norouzi}, \bibinfo{person}{Simone Opel}, \bibinfo{person}{Vee Pettit}, \bibinfo{person}{Leo Porter}, \bibinfo{person}{Brent~N. Reeves}, \bibinfo{person}{Jaromir Savelka}, \bibinfo{person}{IV DavidH.Smith}, \bibinfo{person}{Sven Strickroth}, {and} \bibinfo{person}{Daniel Zingaro}.} \bibinfo{year}{2024}\natexlab{}.
\newblock \showarticletitle{Beyond the Hype: A Comprehensive Review of Current Trends in Generative AI Research, Teaching Practices, and Tools}.
\newblock \bibinfo{journal}{\emph{2024 Working Group Reports on Innovation and Technology in Computer Science Education}} (\bibinfo{year}{2024}).
\newblock


\bibitem[Pu et~al\mbox{.}(2025)]%
        {pu2025can}
\bibfield{author}{\bibinfo{person}{Isabella Pu}, \bibinfo{person}{Prerna Ravi}, \bibinfo{person}{Linh~Dieu Dinh}, \bibinfo{person}{Chelsea Joe}, \bibinfo{person}{Caitlin Ogoe}, \bibinfo{person}{Zixuan Li}, \bibinfo{person}{Cynthia Breazeal}, {and} \bibinfo{person}{Anastasia~K Ostrowski}.} \bibinfo{year}{2025}\natexlab{}.
\newblock \showarticletitle{" How can we learn and use AI at the same time?": Participatory Design of GenAI with High School Students}.
\newblock In \bibinfo{booktitle}{\emph{Proceedings of the 24th Interaction Design and Children}}. \bibinfo{pages}{204--220}.
\newblock


\bibitem[Rafalow(2020)]%
        {rafalow2020digital}
\bibfield{author}{\bibinfo{person}{Matthew~H Rafalow}.} \bibinfo{year}{2020}\natexlab{}.
\newblock \showarticletitle{Digital divisions: How schools create inequality in the tech era}.
\newblock In \bibinfo{booktitle}{\emph{Digital Divisions}}. \bibinfo{publisher}{University of Chicago Press}.
\newblock


\bibitem[Ravi et~al\mbox{.}(2025)]%
        {ravi2025co}
\bibfield{author}{\bibinfo{person}{Prerna Ravi}, \bibinfo{person}{John Masla}, \bibinfo{person}{Gisella Kakoti}, \bibinfo{person}{Grace~C Lin}, \bibinfo{person}{Emma Anderson}, \bibinfo{person}{Matt Taylor}, \bibinfo{person}{Anastasia~K Ostrowski}, \bibinfo{person}{Cynthia Breazeal}, \bibinfo{person}{Eric Klopfer}, {and} \bibinfo{person}{Hal Abelson}.} \bibinfo{year}{2025}\natexlab{}.
\newblock \showarticletitle{Co-designing Large Language Model Tools for Project-Based Learning with K12 Educators}. In \bibinfo{booktitle}{\emph{Proceedings of the 2025 CHI Conference on Human Factors in Computing Systems}}. \bibinfo{pages}{1--25}.
\newblock


\bibitem[Reich(2020)]%
        {reich2020failure}
\bibfield{author}{\bibinfo{person}{Justin Reich}.} \bibinfo{year}{2020}\natexlab{}.
\newblock \bibinfo{booktitle}{\emph{Failure to disrupt: Why technology alone can’t transform education}}.
\newblock \bibinfo{publisher}{Harvard University Press}.
\newblock


\bibitem[Reich et~al\mbox{.}(2017)]%
        {reich2017good}
\bibfield{author}{\bibinfo{person}{Justin Reich}, \bibinfo{person}{Mizuko Ito}, {and} \bibinfo{person}{MS Team}.} \bibinfo{year}{2017}\natexlab{}.
\newblock \showarticletitle{From good intentions to real outcomes}.
\newblock \bibinfo{journal}{\emph{Digital Media and Learning Research Hub. https://clalliance. org/publications/good-intentions-real-outcomes-equity-designlearning-technologies}} (\bibinfo{year}{2017}).
\newblock


\bibitem[Reich et~al\mbox{.}(2012)]%
        {reich2012state}
\bibfield{author}{\bibinfo{person}{Justin Reich}, \bibinfo{person}{Richard Murnane}, {and} \bibinfo{person}{John Willett}.} \bibinfo{year}{2012}\natexlab{}.
\newblock \showarticletitle{The state of wiki usage in US K--12 schools: Leveraging Web 2.0 data warehouses to assess quality and equity in online learning environments}.
\newblock \bibinfo{journal}{\emph{Educational researcher}} \bibinfo{volume}{41}, \bibinfo{number}{1} (\bibinfo{year}{2012}), \bibinfo{pages}{7--15}.
\newblock


\bibitem[Roque(2023)]%
        {roque2023imagining}
\bibfield{author}{\bibinfo{person}{Ricarose Roque}.} \bibinfo{year}{2023}\natexlab{}.
\newblock \showarticletitle{Imagining Alternative Visions of Computing: Photo-Visuals of Material, Social, and Emotional Contexts from Family Creative Learning}. In \bibinfo{booktitle}{\emph{Proceedings of the 22nd Annual ACM Interaction Design and Children Conference}} (Chicago, IL, USA) \emph{(\bibinfo{series}{IDC '23})}. \bibinfo{publisher}{Association for Computing Machinery}, \bibinfo{address}{New York, NY, USA}, \bibinfo{pages}{68--81}.
\newblock
\showISBNx{9798400701096}
\urldef\tempurl%
\url{https://doi.org/10.1145/3585088.3589364}
\showDOI{\tempurl}


\bibitem[Schneider et~al\mbox{.}(2025)]%
        {schneider2025thematic}
\bibfield{author}{\bibinfo{person}{Johannes Schneider}, \bibinfo{person}{B{\'e}atrice~S Hasler}, \bibinfo{person}{Michaela Varrone}, \bibinfo{person}{Fabian Hoya}, \bibinfo{person}{Thomas Schroffenegger}, \bibinfo{person}{Dana-Kristin Mah}, {and} \bibinfo{person}{Karl Peb{\"o}ck}.} \bibinfo{year}{2025}\natexlab{}.
\newblock \showarticletitle{Thematic and Task-Based Categorization of K-12 GenAI Usages with Hierarchical Topic Modeling}.
\newblock \bibinfo{journal}{\emph{arXiv preprint arXiv:2508.09997}} (\bibinfo{year}{2025}).
\newblock


\bibitem[Schoenfeld(2015)]%
        {Schoenfeld2015HowWT}
\bibfield{author}{\bibinfo{person}{Alan~H. Schoenfeld}.} \bibinfo{year}{2015}\natexlab{}.
\newblock \showarticletitle{How We Think: A Theory of Human Decision-Making, with a Focus on Teaching}.
\newblock
\urldef\tempurl%
\url{https://api.semanticscholar.org/CorpusID:150680880}
\showURL{%
\tempurl}


\bibitem[Schuck and Kearney(2008)]%
        {Schuck2008ClassroomBasedUO}
\bibfield{author}{\bibinfo{person}{Sandy Schuck} {and} \bibinfo{person}{Matthew Kearney}.} \bibinfo{year}{2008}\natexlab{}.
\newblock \showarticletitle{Classroom-Based Use of Two Educational Technologies: A Sociocultural Perspective}.
\newblock \bibinfo{journal}{\emph{Contemporary Issues in Technology and Teacher Education}}  \bibinfo{volume}{8} (\bibinfo{year}{2008}), \bibinfo{pages}{394--406}.
\newblock
\urldef\tempurl%
\url{https://api.semanticscholar.org/CorpusID:18216849}
\showURL{%
\tempurl}


\bibitem[Selwyn(2013)]%
        {Selwyn2013EmpoweringTW}
\bibfield{author}{\bibinfo{person}{Neil Selwyn}.} \bibinfo{year}{2013}\natexlab{}.
\newblock \showarticletitle{“Empowering the World’s Poorest Children”? A Critical Examination of One Laptop per Child}.
\newblock
\urldef\tempurl%
\url{https://api.semanticscholar.org/CorpusID:114668574}
\showURL{%
\tempurl}


\bibitem[Selwyn et~al\mbox{.}(2016)]%
        {selwyn2016high}
\bibfield{author}{\bibinfo{person}{Neil Selwyn}, \bibinfo{person}{Selena Nemorin}, {and} \bibinfo{person}{Nicola Johnson}.} \bibinfo{year}{2016}\natexlab{}.
\newblock \showarticletitle{High-tech, hard work: an investigation of teachers’ work in the digital age}.
\newblock \bibinfo{journal}{\emph{Learning, Media and Technology}} \bibinfo{volume}{42}, \bibinfo{number}{4} (\bibinfo{year}{2016}), \bibinfo{pages}{390--405}.
\newblock
\urldef\tempurl%
\url{https://doi.org/10.1080/17439884.2016.1252770}
\showDOI{\tempurl}


\bibitem[Shen et~al\mbox{.}(2024)]%
        {shen2024implications}
\bibfield{author}{\bibinfo{person}{Yiyin Shen}, \bibinfo{person}{Xinyi Ai}, \bibinfo{person}{Adalbert~Gerald Soosai~Raj}, \bibinfo{person}{Rogers~Jeffrey Leo~John}, {and} \bibinfo{person}{Meenakshi Syamkumar}.} \bibinfo{year}{2024}\natexlab{}.
\newblock \showarticletitle{Implications of chatgpt for data science education}. In \bibinfo{booktitle}{\emph{Proceedings of the 55th ACM Technical Symposium on Computer Science Education V. 1}}. \bibinfo{pages}{1230--1236}.
\newblock


\bibitem[Shengjergji et~al\mbox{.}(2024)]%
        {Shengjergji2024EnvironmentalIO}
\bibfield{author}{\bibinfo{person}{Sofije Shengjergji}, \bibinfo{person}{Anna Luzai}, \bibinfo{person}{Stephanie Mills}, \bibinfo{person}{Parker~Van Nostrand}, \bibinfo{person}{Anna~Lindroos Cermakova}, {and} \bibinfo{person}{Natalia~Ingebretsen Kucirkova}.} \bibinfo{year}{2024}\natexlab{}.
\newblock \showarticletitle{Environmental impact of EdTech: The hidden costs of digital learing}.
\newblock


\bibitem[Simon and Johnson(2015)]%
        {simon2015teacher}
\bibfield{author}{\bibinfo{person}{Nicole Simon} {and} \bibinfo{person}{Susan~Moore Johnson}.} \bibinfo{year}{2015}\natexlab{}.
\newblock \showarticletitle{Teacher turnover in high-poverty schools: What we know and can do}.
\newblock \bibinfo{journal}{\emph{Teachers College Record}} \bibinfo{volume}{117}, \bibinfo{number}{3} (\bibinfo{year}{2015}), \bibinfo{pages}{1--36}.
\newblock


\bibitem[Singer(2025a)]%
        {nyt2025miamidade}
\bibfield{author}{\bibinfo{person}{Natasha Singer}.} \bibinfo{year}{2025}\natexlab{a}.
\newblock \bibinfo{title}{How Miami Schools Are Leading 100,000 Students Into the A.I. Future}.
\newblock \bibinfo{howpublished}{\url{https://www.nytimes.com/2025/05/19/technology/ai-miami-schools-google-gemini.html}}.
\newblock
\newblock
\shownote{[Accessed 07-29-2025]}.


\bibitem[Singer(2025b)]%
        {nyt2025microsoft}
\bibfield{author}{\bibinfo{person}{Natasha Singer}.} \bibinfo{year}{2025}\natexlab{b}.
\newblock \bibinfo{title}{Microsoft Pledges \$4 Billion Toward A.I. Education}.
\newblock \bibinfo{howpublished}{\url{https://www.nytimes.com/2025/07/09/business/microsoft-ai-education.html}}.
\newblock
\newblock
\shownote{Accessed 07-29-2025}.


\bibitem[Singer(2025c)]%
        {nyt2025bankroll}
\bibfield{author}{\bibinfo{person}{Natasha Singer}.} \bibinfo{year}{2025}\natexlab{c}.
\newblock \bibinfo{title}{OpenAI and Microsoft Bankroll New A.I. Training for Teachers}.
\newblock \bibinfo{howpublished}{\url{https://www.nytimes.com/2025/07/08/technology/chatgpt-teachers-openai-microsoft.html}}.
\newblock
\newblock
\shownote{Accessed 07-29-2025}.


\bibitem[Solyst et~al\mbox{.}(2023)]%
        {solyst2023would}
\bibfield{author}{\bibinfo{person}{Jaemarie Solyst}, \bibinfo{person}{Shixian Xie}, \bibinfo{person}{Ellia Yang}, \bibinfo{person}{Angela~EB Stewart}, \bibinfo{person}{Motahhare Eslami}, \bibinfo{person}{Jessica Hammer}, {and} \bibinfo{person}{Amy Ogan}.} \bibinfo{year}{2023}\natexlab{}.
\newblock \showarticletitle{“I Would Like to Design”: Black Girls Analyzing and Ideating Fair and Accountable AI}. In \bibinfo{booktitle}{\emph{Proceedings of the 2023 CHI Conference on Human Factors in Computing Systems}}. \bibinfo{pages}{1--14}.
\newblock


\bibitem[Sommerfeld(2020)]%
        {Sommerfeld2020AmesMG}
\bibfield{author}{\bibinfo{person}{Alice Sommerfeld}.} \bibinfo{year}{2020}\natexlab{}.
\newblock \showarticletitle{Ames, Morgan G.: The Charisma Machine. The Life, Death, and Legacy of One Laptop per Child. Cambridge/Massachusetts 2019.}
\newblock \bibinfo{journal}{\emph{Rhetorik}}  \bibinfo{volume}{39} (\bibinfo{year}{2020}), \bibinfo{pages}{107 -- 111}.
\newblock


\bibitem[Song et~al\mbox{.}(2024)]%
        {song2024students}
\bibfield{author}{\bibinfo{person}{Yukyeong Song}, \bibinfo{person}{Jinhee Kim}, \bibinfo{person}{Zifeng Liu}, \bibinfo{person}{Chenglu Li}, {and} \bibinfo{person}{Wanli Xing}.} \bibinfo{year}{2024}\natexlab{}.
\newblock \showarticletitle{Students’ perceived roles, opportunities, and challenges of a generative AI-powered teachable agent: a case of middle school math class}.
\newblock \bibinfo{journal}{\emph{Journal of Research on Technology in Education}} (\bibinfo{year}{2024}), \bibinfo{pages}{1--19}.
\newblock


\bibitem[Suh et~al\mbox{.}(2024)]%
        {suh2024opportunities}
\bibfield{author}{\bibinfo{person}{Hyewon Suh}, \bibinfo{person}{Aayushi Dangol}, \bibinfo{person}{Hedda Meadan}, \bibinfo{person}{Carol~A Miller}, {and} \bibinfo{person}{Julie~A Kientz}.} \bibinfo{year}{2024}\natexlab{}.
\newblock \showarticletitle{Opportunities and challenges for AI-based support for speech-language pathologists}. In \bibinfo{booktitle}{\emph{Proceedings of the 3rd Annual Meeting of the Symposium on Human-Computer Interaction for Work}}. \bibinfo{pages}{1--14}.
\newblock


\bibitem[Tafazoli(2024)]%
        {tafazoli2024exploring}
\bibfield{author}{\bibinfo{person}{Dara Tafazoli}.} \bibinfo{year}{2024}\natexlab{}.
\newblock \showarticletitle{Exploring the potential of generative AI in democratizing English language education}.
\newblock \bibinfo{journal}{\emph{Computers and Education: Artificial Intelligence}}  \bibinfo{volume}{7} (\bibinfo{year}{2024}), \bibinfo{pages}{100275}.
\newblock


\bibitem[Toyama(2015)]%
        {toyama2015geek}
\bibfield{author}{\bibinfo{person}{Kentaro Toyama}.} \bibinfo{year}{2015}\natexlab{}.
\newblock \bibinfo{booktitle}{\emph{Geek heresy: Rescuing social change from the cult of technology}}.
\newblock \bibinfo{publisher}{PublicAffairs}.
\newblock


\bibitem[Valdivieso and Gonz{\'a}lez(2025)]%
        {Valdivieso2025GenerativeAT}
\bibfield{author}{\bibinfo{person}{Tizziana Valdivieso} {and} \bibinfo{person}{Oscar Gonz{\'a}lez}.} \bibinfo{year}{2025}\natexlab{}.
\newblock \showarticletitle{Generative AI Tools in Salvadoran Higher Education: Balancing Equity, Ethics, and Knowledge Management in the Global South}.
\newblock \bibinfo{journal}{\emph{Education Sciences}} (\bibinfo{year}{2025}).
\newblock


\bibitem[Varanasi et~al\mbox{.}(2019)]%
        {Varanasi2019HowTI}
\bibfield{author}{\bibinfo{person}{Rama~Adithya Varanasi}, \bibinfo{person}{Ren{\'e}~F. Kizilcec}, {and} \bibinfo{person}{Nicola Dell}.} \bibinfo{year}{2019}\natexlab{}.
\newblock \showarticletitle{How Teachers in India Reconfigure their Work Practices around a Teacher-Oriented Technology Intervention}.
\newblock \bibinfo{journal}{\emph{Proceedings of the ACM on Human-Computer Interaction}}  \bibinfo{volume}{3} (\bibinfo{year}{2019}), \bibinfo{pages}{1 -- 21}.
\newblock


\bibitem[Walter(2024)]%
        {Walter2024EmbracingTF}
\bibfield{author}{\bibinfo{person}{Yoshija Walter}.} \bibinfo{year}{2024}\natexlab{}.
\newblock \showarticletitle{Embracing the future of Artificial Intelligence in the classroom: the relevance of AI literacy, prompt engineering, and critical thinking in modern education}.
\newblock \bibinfo{journal}{\emph{International Journal of Educational Technology in Higher Education}}  \bibinfo{volume}{21} (\bibinfo{year}{2024}), \bibinfo{pages}{1--29}.
\newblock


\bibitem[Wang(2025)]%
        {Wang2025ScaffoldingCI}
\bibfield{author}{\bibinfo{person}{Nicole~C. Wang}.} \bibinfo{year}{2025}\natexlab{}.
\newblock \showarticletitle{Scaffolding Creativity: Integrating Generative AI Tools and Real-World Experiences in Business Education}.
\newblock \bibinfo{journal}{\emph{Proceedings of the Extended Abstracts of the CHI Conference on Human Factors in Computing Systems}} (\bibinfo{year}{2025}).
\newblock
\urldef\tempurl%
\url{https://api.semanticscholar.org/CorpusID:275470801}
\showURL{%
\tempurl}


\bibitem[Wang and Baker(2015)]%
        {wang2015content}
\bibfield{author}{\bibinfo{person}{Yuan Wang} {and} \bibinfo{person}{Ryan Baker}.} \bibinfo{year}{2015}\natexlab{}.
\newblock \showarticletitle{Content or platform: Why do students complete MOOCs}.
\newblock \bibinfo{journal}{\emph{MERLOT Journal of Online Learning and Teaching}} \bibinfo{volume}{11}, \bibinfo{number}{1} (\bibinfo{year}{2015}), \bibinfo{pages}{17--30}.
\newblock


\bibitem[Willet and He(2024)]%
        {StaudtWillet2024EducatorsIL}
\bibfield{author}{\bibinfo{person}{K.~Bret~Staudt Willet} {and} \bibinfo{person}{Dan He}.} \bibinfo{year}{2024}\natexlab{}.
\newblock \showarticletitle{Educators' invisible labour: A systematic review}.
\newblock \bibinfo{journal}{\emph{Review of Education}} (\bibinfo{year}{2024}).
\newblock


\bibitem[Williams et~al\mbox{.}(2019)]%
        {williams_is_2019}
\bibfield{author}{\bibinfo{person}{Randi Williams}, \bibinfo{person}{Hae~Won Park}, {and} \bibinfo{person}{Cynthia Breazeal}.} \bibinfo{year}{2019}\natexlab{}.
\newblock \showarticletitle{A is for {Artificial} {Intelligence}: {The} {Impact} of {Artificial} {Intelligence} {Activities} on {Young} {Children}'s {Perceptions} of {Robots}}. In \bibinfo{booktitle}{\emph{Proceedings of the 2019 {CHI} {Conference} on {Human} {Factors} in {Computing} {Systems}}} \emph{(\bibinfo{series}{{CHI} '19})}. \bibinfo{publisher}{Association for Computing Machinery}, \bibinfo{address}{New York, NY, USA}, \bibinfo{pages}{1--11}.
\newblock
\showISBNx{978-1-4503-5970-2}
\urldef\tempurl%
\url{https://doi.org/10.1145/3290605.3300677}
\showDOI{\tempurl}


\bibitem[Williamson(2024)]%
        {williamson2024ai}
\bibfield{author}{\bibinfo{person}{Ben Williamson}.} \bibinfo{year}{2024}\natexlab{}.
\newblock \showarticletitle{AI in education is a public problem}.
\newblock \bibinfo{journal}{\emph{Code Acts in Education}} (\bibinfo{year}{2024}).
\newblock


\bibitem[Windschitl and Sahl(2002)]%
        {Windschitl2002TracingTU}
\bibfield{author}{\bibinfo{person}{Mark Windschitl} {and} \bibinfo{person}{Kurt Sahl}.} \bibinfo{year}{2002}\natexlab{}.
\newblock \showarticletitle{Tracing Teachers’ Use of Technology in a Laptop Computer School: The Interplay of Teacher Beliefs, Social Dynamics, and Institutional Culture}.
\newblock \bibinfo{journal}{\emph{American Educational Research Journal}}  \bibinfo{volume}{39} (\bibinfo{year}{2002}), \bibinfo{pages}{165 -- 205}.
\newblock


\bibitem[Xie et~al\mbox{.}(2024)]%
        {Xie2024CodesigningAE}
\bibfield{author}{\bibinfo{person}{Benjamin Xie}, \bibinfo{person}{Parth Sarin}, \bibinfo{person}{Jacob Wolf}, \bibinfo{person}{Raycelle C.~C. Garcia}, \bibinfo{person}{Victoria Delaney}, \bibinfo{person}{Isabel Sieh}, \bibinfo{person}{Anika Fuloria}, \bibinfo{person}{Deepak~Varuvel Dennison}, \bibinfo{person}{Christine Bywater}, {and} \bibinfo{person}{Victor~R. Lee}.} \bibinfo{year}{2024}\natexlab{}.
\newblock \showarticletitle{Co-designing AI Education Curriculum with Cross-Disciplinary High School Teachers}. In \bibinfo{booktitle}{\emph{AAAI Conference on Artificial Intelligence}}.
\newblock


\bibitem[Yang et~al\mbox{.}(2024)]%
        {Yang2024DesigningAA}
\bibfield{author}{\bibinfo{person}{Ellia Yang}, \bibinfo{person}{Amy Ogan}, \bibinfo{person}{Jessica Hammer}, {and} \bibinfo{person}{Jaemarie Solyst}.} \bibinfo{year}{2024}\natexlab{}.
\newblock \showarticletitle{Designing an AI Literacy Transformational Game for Families}.
\newblock \bibinfo{journal}{\emph{Proceedings of the 2024 ACM Conference on International Computing Education Research - Volume 2}} (\bibinfo{year}{2024}).
\newblock


\bibitem[Ye et~al\mbox{.}(2025)]%
        {Ye2025TheDS}
\bibfield{author}{\bibinfo{person}{Runlong Ye}, \bibinfo{person}{Matthew Varona}, \bibinfo{person}{Oliver Huang}, \bibinfo{person}{Patrick Yung~Kang Lee}, \bibinfo{person}{Michael Liut}, {and} \bibinfo{person}{Carolina Nobre}.} \bibinfo{year}{2025}\natexlab{}.
\newblock \showarticletitle{The Design Space of Recent AI-assisted Research Tools for Ideation, Sensemaking, and Scientific Creativity}.
\newblock \bibinfo{journal}{\emph{ArXiv}}  \bibinfo{volume}{abs/2502.16291} (\bibinfo{year}{2025}).
\newblock
\urldef\tempurl%
\url{https://api.semanticscholar.org/CorpusID:276574980}
\showURL{%
\tempurl}


\bibitem[Yen and Hsu(2023)]%
        {Yen2023ThreeQC}
\bibfield{author}{\bibinfo{person}{An-Zi Yen} {and} \bibinfo{person}{Wei-Ling Hsu}.} \bibinfo{year}{2023}\natexlab{}.
\newblock \showarticletitle{Three Questions Concerning the Use of Large Language Models to Facilitate Mathematics Learning}. In \bibinfo{booktitle}{\emph{Conference on Empirical Methods in Natural Language Processing}}.
\newblock
\urldef\tempurl%
\url{https://api.semanticscholar.org/CorpusID:264405766}
\showURL{%
\tempurl}


\bibitem[Yin et~al\mbox{.}(2025)]%
        {Yin2025ResponsibleAI}
\bibfield{author}{\bibinfo{person}{Yaxuan Yin}, \bibinfo{person}{Shamya Karumbaiah}, {and} \bibinfo{person}{Shona Acquaye}.} \bibinfo{year}{2025}\natexlab{}.
\newblock \showarticletitle{Responsible AI in Education: Understanding Teachers’ Priorities and Contextual Challenges}.
\newblock \bibinfo{journal}{\emph{Proceedings of the 2025 ACM Conference on Fairness, Accountability, and Transparency}} (\bibinfo{year}{2025}).
\newblock


\bibitem[Zheng et~al\mbox{.}(2025)]%
        {zheng2025students}
\bibfield{author}{\bibinfo{person}{Jiayu Zheng}, \bibinfo{person}{Lingxin Hao}, \bibinfo{person}{Kelun Lu}, \bibinfo{person}{Ashi Garg}, \bibinfo{person}{Mike Reese}, \bibinfo{person}{Melo-Jean Yap}, \bibinfo{person}{I Wang}, \bibinfo{person}{Xingyun Wu}, \bibinfo{person}{Wenrui Huang}, \bibinfo{person}{Jenna Hoffman}, {et~al\mbox{.}}} \bibinfo{year}{2025}\natexlab{}.
\newblock \showarticletitle{Do Students Rely on AI? Analysis of Student-ChatGPT Conversations from a Field Study}.
\newblock \bibinfo{journal}{\emph{arXiv preprint arXiv:2508.20244}} (\bibinfo{year}{2025}).
\newblock


\end{thebibliography}

\end{document}